\documentclass[reprint,pre,aps,floatfix]{revtex4-2}
\usepackage{graphicx}
\usepackage{amsfonts}
\usepackage{amssymb}
\usepackage{natbib}
\usepackage{xcolor}
\usepackage{amsmath}
\usepackage{enumerate}
\usepackage[percent]{overpic}
\usepackage{verbatim}
\usepackage[colorlinks=true, linkcolor=blue, citecolor=blue, urlcolor=blue]{hyperref}
\usepackage{booktabs}
\usepackage{siunitx}

\usepackage{environ}
\NewEnviron{subfigure}[2][]{%
  \begin{minipage}[#1]{#2}\BODY\end{minipage}%
}

\usepackage{ragged2e}
\makeatletter
\long\def\@makecaption#1#2{%
  \vskip\abovecaptionskip
  \noindent\justifying
  \hangindent=0.0em\hangafter=1 
  \leavevmode\normalfont#1\nobreak\hskip.5em\relax #2\par
  \vskip\belowcaptionskip}
\makeatother

\newcommand{\be}{\begin{equation}}
\newcommand{\ee}{\end{equation}}
\newcommand{\bea}{\begin{eqnarray}}
\newcommand{\eea}{\end{eqnarray}}

\newcommand{\curlyL}{$\mathop{\mathcal{L}}$}

\begin{document}
\title{Quantum chaos and semiclassical behavior in mushroom billiards II: Structure of quantum eigenstates and their phase space localization properties}

\author{Matic Orel and Marko Robnik}

\affiliation{CAMTP - Center for Applied Mathematics and Theoretical
  Physics, University of Maribor, Mladinska 3, SI-2000 Maribor, Slovenia,
  European Union}

\date{\today}

\begin{abstract}
We investigate eigenstate localization in the phase space of the Bunimovich mushroom billiard, a paradigmatic mixed–phase–space system whose piecewise-$C^1$ boundary yields a single clean separatrix between one regular and one chaotic region. By varying the stem half-width $w$, we continuously change the strength and extent of bouncing-ball stickiness in the stem, which for narrow stems gives rise to phase space localization of chaotic eigenstates. Using the Poincaré–Husimi (PH) representation of eigenstates we quantify localization via information entropies and inverse participation ratios of PH functions. For sufficiently wide stems the distribution of entropy localization measures converges to a two-parameter beta distribution, while entropy localization measures and inverse participation ratios across the chaotic ensemble exhibit an approximately linear relationship. Finally, the fraction of mixed (neither purely regular nor fully chaotic) eigenstates decays as a power-law in the effective semiclassical parameter, in precise agreement with the Principle of Uniform Semiclassical Condensation of Wigner functions (PUSC)\cite{Rob1998,Rob2023A}.
\end{abstract}

\maketitle

\section{Introduction}
\label{introduction}

Quantum chaos (or more generally, wave chaos) investigates how classically chaotic dynamics imprint themselves on quantum or wave systems \cite{Stoe,Haake}.  A remarkable correspondence emerges: classically integrable systems exhibit uncorrelated (Poisson) energy‐level statistics \cite{BerTab1977,Veble_Robnik_integrable}, whereas fully chaotic (ergodic) systems display the level‐repulsion characteristic of random‐matrix theory (GOE or GUE), as conjectured by Bohigas, Giannoni and Schmit \cite{BGS1984,Cas1980}.  Beyond spectral tests, the phase–space structure of eigenstates likewise reflects classical invariants: Wigner and Poincar\'e-Husimi distributions condense on invariant tori for regular motion, or in the semiclassical limit uniformly fill the chaotic sea for ergodic motion. Mixed‐type systems—whose classical phase space is partitioned into coexisting regular islands and a chaotic sea—exhibit a wealth of nontrivial quantum signatures (e.g. tunneling \cite{Vidmar2007}).  According to the Berry–Robnik picture, in the strict semiclassical limit ($\hbar\to0$) the energy spectrum decomposes into statistically independent subsequences: one arising from states supported on invariant tori (regular levels, Poisson statistics), and others from states filling the chaotic regions (chaotic levels, random‐matrix statistics) \cite{BerRob1984}. Correspondingly, individual eigenfunctions concentrate either on invariant tori or on the chaotic regions, in accordance with Percival’s conjecture and PUSC \cite{Percival1973,BerRob1984,Rob2023,Rob1998}. In practice, however, there is an interplay of two important time scales.  The Ehrenfest time  
\[
  t_E = \frac{1}{\lambda}\,\ln\!\bigl(\frac{1}{\hbar}\bigr),
\]
marks the duration over which a minimal wavepacket follows the underlying classical dynamics (with Lyapunov exponent $\lambda$), while the Heisenberg time  
\[
  t_H = \frac{\hbar}{\Delta E},
\]
is set by the inverse mean level spacing $\Delta E$. Another important time scale in the semiclassical limit is the classical transport time $t_T$. The ratio 
\[
  \alpha = \frac{t_H}{t_T}
\]
governs the onset of quantum ergodicity in the chaotic sector: for $\alpha\gg1$, classical diffusion \cite{Walters1982} has time to fully develop before quantum interference occurs, so chaotic eigenstates can fully spread; but when $\alpha\lesssim1$, interferences due to spectrum discretization give rise to dynamical localization. Even in a classically fully chaotic region, at any finite $\hbar$ some eigenstates fail to occupy the full ergodic phase‐space volume and are localized instead. Also they often exhibit enhanced intensity around remnants of classical barriers like the cantori in smooth Hamiltonians \cite{MacKayMeissPercival_transport_hamiltonian_systems} or marginally unstable periodic orbits (MUPOs) in billiards—so that their Husimi (or Wigner) distributions show enhanced density near these sticky structures \cite{Lozej2020,crt_stickiness}. This intermediate regime of localized chaotic states is characterized by deviations from both Poisson and Random Matrix Theory (RMT) statistics in level spacings, often captured by phenomenological distributions (Brody \cite{Bro1973}, Izrailev \cite{Izr1989}) which interpolate between the two limits.  

Billiard systems—point particles confined to a planar domain \(B\subset\mathbb{R}^2\) with perfectly specular reflections off its boundary $\partial B$ — provide a remarkably rich yet mathematically transparent setting for classical and quantum dynamics. Classically, the motion consists of straight–line motion broken up by instantaneous reflections described by an exact, area‐preserving Poincar\'e map on the surface of a cylinder (the bounce map). The curvature of \(\partial B\) governs the dynamical regime \cite{Steiner1994}:
\begin{itemize}
  \item \emph{Integrable billiards} (circle, ellipse, rectangle) admit continuous families of invariant tori and show no chaos.
  \item \emph{Dispersing billiards \cite{Sinai1970}} (e.g. Sinai billiard) with strictly convex scatterers are uniformly hyperbolic.
  \item \emph{Defocusing billiards \cite{Bun1979}} (e.g. stadium) combine curved arcs with flat segments to yield ergodicity via the defocusing mechanism of Bunimovich \cite{Bunimovich2014}.
\end{itemize}

In general, billiards are of the mixed-type with regular and chaotic regions coexisting in the phase space. Additional hallmark phenomena include MUPOs that can cause stickiness \cite{Altmann2005}, whispering‐gallery modes along convex walls \cite{Lazutkin_1973}, and nontrivial transport barriers formed by e.g. cantori. This blend of geometric simplicity and dynamical complexity makes billiards a cornerstone for testing semiclassical theories, quantum spectral statistics, and wave–chaos phenomena in mixed‐phase‐space systems. 

\begin{figure}
    \includegraphics[width=1.0\linewidth]{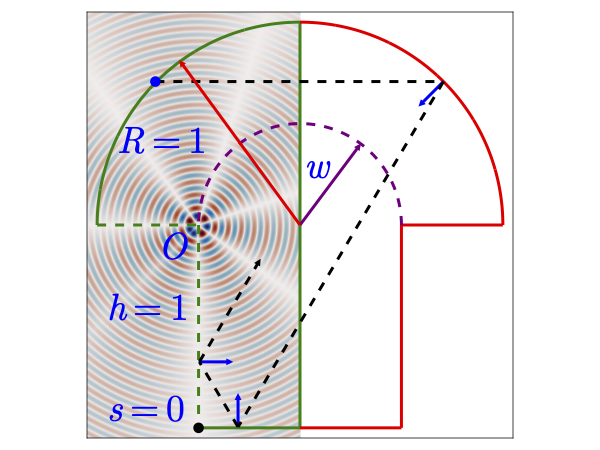}
    \caption{Boundary representing the mushroom billiard. For the classical dynamics we use the full boundary and for the quantum calculations the half mushroom boundary represented by the green boundary. The Poincar\'e-Birkhoff coordinates are $(s,p)$, where $s$ is counted anti-clockwise from the origin $s=0$, while $p$ is the sine of the reflection angle. The origin in configuration space is designated by $O=(0,0)$. It coincides with the re-entrant corner for technical reasons \protect\cite{betcke}. On the dashed green vertical lines the Dirichlet BC are automatically satisfied by the construction (CAFB). The purple colored dashed curve is the inscribed circle to which MUPOs forming the boundary between the regular and chaotic region must be tangent to.}
    \label{geom_mushroom}
\end{figure}

The Bunimovich “mushroom’’ billiard \cite{Bunimovich2001}(see Fig.\ref{geom_mushroom}) is a particularly clean mixed‐type example: a circular cap of radius $R$ joined to a rectangular-stem of half–width $w<R$ and length $h$ yields exactly one regular component (cap-trapped trajectories) and one chaotic component (all others), separated by a single, analytically known separatrix. In this work we use the flexible geometry of Bunimovich’s mushroom billiard by varying \(w\) to probe localization of quantum eigenstates in a mixed phase space setting. Using Poincaré–Husimi (PH) functions \cite{Hus1940,TV1995,Backer2003} as a positive‐definite phase–space representation, we introduce two complementary measures of localization, one based on information entropy \(A_n\) \cite{BatRob2013A}, and the inverse–participation ratio \(\mathrm{IPR}_n\) \cite{Imre_Varga_1998_IPR}. We employ them both to characterize localization properties of regular, chaotic and mixed type eigenstates. By systematically scanning over \(w\) and wavenumber \(k\), we investigate (i) the variation of the full chaotic‐state distribution \(P(A)\), (ii) the joint behavior of \(\{A_n,\mathrm{IPR}_n\}\), and (iii) the decay of the mixed‐state fraction \(\chi\) in the semiclassical parameter, for which we choose the leading term of Weyl's law \cite{Weyl1911}

\begin{equation} \label{semiclassical_parameter}
    e \;=\;\frac{\mathcal{A}}{4\pi}\,k^2\,,
\end{equation}

\noindent where $\mathcal{A}$ is the area of the billiard. Such an approach allows a detailed analysis of the PUSC in a setting with a sharply divided but tunable phase space.  

The paper is organized as follows.  In Sec.~\ref{geometry} we define the mushroom geometry and its classical phase space along with defining the quantum problem and the technique of solving it.  Sec.~\ref{husimi} introduces the Poincar\'e-Husimi functions. Sec.~\ref{separation} introduces the overlap $M$ index and its use in separating the chaotic and regular states. Sec.~\ref{localizations} defines the entropy localization measure and inverse participation ratio along with numerical results for their distributions. Sec.~\ref{mixed_states} analyzes the mixed-state fraction and its power-law decay. We present our conclusions and perspectives in Sec.~\ref{conclusion}.

\begin{figure}[ht]
  \begin{subfigure}[t]{0.5\linewidth}
    \begin{overpic}[width=\linewidth]{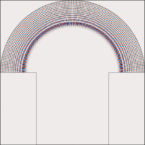}
      \put(3,90){\large (a)}
    \end{overpic}
  \end{subfigure}\hfill
  \begin{subfigure}[t]{0.5\linewidth}
    \begin{overpic}[width=\linewidth]{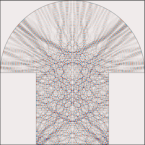}
      \put(3,90){\large\ (b)}
    \end{overpic}
  \end{subfigure}

  \begin{subfigure}[t]{0.5\linewidth}
    \begin{overpic}[width=\linewidth]{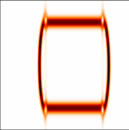}
      \put(3,90){\large (c)}
    \end{overpic}
  \end{subfigure}\hfill
  \begin{subfigure}[t]{0.5\linewidth}
    \begin{overpic}[width=\linewidth]{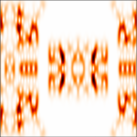}
      \put(3,90){\large (d)}
    \end{overpic}
  \end{subfigure}

  \caption{
    (Top row) Typical wavefunction plots for a (a) regular state ($k=213.949$) and (b) a non-localized chaotic state ($k=214.008$) for $w=0.5$. (Bottom row) (c) Poincar\'e-Husimi function of the regular wavefunction above, (d) Poincar\'e-Husimi function of the chaotic wavefunction above.}
    \label{representative_states}
\end{figure}

\section{Mushroom Billiard Geometry}
\label{geometry}

The mushroom billiard consists of a semicircular “cap” (a focusing boundary segment)\cite{Bunimovich2001} of radius $R=1$ attached to a rectangular “stem” of height $h=1$ and total width $2w$, with $w$ ranging from $0.1$ to $0.9$. As we are interested in the mixed-type regime, we study here the cases $w=0.1-0.9$ \cite{OreLozRobYan2025}. In the extreme case $w=0$ we have the integrable case of the circle, while $w=1$ is the case of the fully chaotic Bunimovich stadium. In our classical calculations we employ the full, symmetric mushroom, whereas all quantum calculations use the desymmetrized half–billiard (odd symmetry class) with Dirichlet boundary conditions along its vertical symmetry axis \cite{BarBet2007}.  Fig.\ref{geom_mushroom} shows both the full (red+green) and desymmetrized (green) boundaries, with the corner-adapted Fourier–Bessel (CAFB) \cite{betcke} basis overlaid on the latter.  For phase–space visualizations we use Poincaré–Birkhoff coordinates $(s,p)$ \cite{chernov2006_Poincare_sections,Berry1981}, where $s$ is the boundary arclength (measured counterclockwise from the marked origin $s=0$) and $p=\sin\theta$ is the conjugate momentum, where $\theta$ is the reflection angle. These variables facilitate a direct comparison between classical trajectories and the Poincar\'e–Husimi projections of quantum eigenstates.

Although each straight or circular segment of the mushroom boundary is at least \(C^1\), the presence of sharp corners reduces the overall smoothness to only \(C^0\).  Hence the system lies outside the usual Kolmogorov–Arnold–Moser (KAM) theory \cite{kolmogorov1954,Arnold1963,moser1962} of near‐integrable Hamiltonians with fractal hierarchies of island chains and sticky cantori \cite{MacKayMeissPercival_transport_hamiltonian_systems}.  Instead, the mushroom billiard admits exactly one regular island and one chaotic sea, separated by a single clean separatrix. Stickiness occurs in two varieties—external and internal \cite{Bunimovich_ext_int_stickiness}. Internal stickiness is generated by MUPOs in the stem, while external stickiness by either MUPOs or sparse points in the chaotic region that are just tangent to the dashed semicircle in Fig.\ref{geom_mushroom}. For any stem half-width \(w>0\) there exist bouncing‐ball MUPOs trapped between the parallel straight walls of the stem in the chaotic region while the same cannot be said for the MUPOs trapped in the cap \cite{altmann_mupo_criterion} (they consist of orbits tangent to the inscribed inner circle of radius $w$). For $R=1$ the minimal half-width of the stem required to potentially admit MUPOs must be $w \gtrapprox 0.390683$ along with meeting the following condition: 

\begin{equation} \label{MUPO_check}
    \sin(\pi \theta_p(q,\eta)) < w/R < \frac{ \sin(\pi \theta_p(q,\eta))}{ \cos(\pi/(jq))}, \quad |\theta_p(q,\eta)|= \frac{q-2\eta}{2q}
\end{equation}

\noindent where $q,\eta \in \mathbb{N}$ are the period and rotation number of an orbit that is a MUPO \cite{Altmann2005}. Detailed checks of \eqref{MUPO_check} are found in Appendix \ref{stickiness_detailed} where we consider the triangular stem mushroom billiard geometry in order to eliminate the role of bouncing-ball modes. 

Quantum analysis requires solving the stationary Schr\"odinger equation (Helmholtz equation)

\begin{equation} \label{helmholtz_eq}
    (\Delta+k^2)\psi_n =0,
\end{equation}

\noindent subject to Dirichlet boundary conditions (BC) $\psi_n|_{\partial B}=0$ where $B$ is the billiard interior domain. For the specifics of the numerical method we refer to our previous paper \cite{OreLozRobYan2025}. Due to the nature of using the Vergini-Saraceno scaling method \cite{VerSar1995} it can happen that some levels could be missing from the spectrum. To quantify our loss of energy levels we compared their number to their expected number as given by Weyl's law

\begin{equation} \label{weyl_law}
    N(k)=\frac{\mathcal{A}}{4\pi}k^2-\frac{\mathcal{L}}{4\pi}k+c.c,
\end{equation}

\noindent where $\mathcal{A}$ is the area of the desymmetrized domain of the mushroom billiard, $\mathcal{L}$ is the length of the boundary of the desymmetrized domain and $c.c$ are corner and curvature corrections which are independent of $k$. For all geometries the fraction of missing levels was $<0.1\%$

\begin{figure}[ht]
  \begin{subfigure}[t]{0.35\linewidth}
    \begin{overpic}[width=\linewidth,height=2.0in]{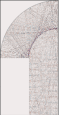}
      \put(3,90){\large (a)}
    \end{overpic}
  \end{subfigure}\hfill
  \begin{subfigure}[t]{0.65\linewidth}
    \begin{overpic}[width=\linewidth,height=2.0in]{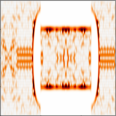}
      \put(2,81){\large\ (b)}
    \end{overpic}
  \end{subfigure}
  \caption{
    Example of a mixed-type state (a) wavefunction and (b) PH function ($M=0.51,w=0.5$) showing increased density around the phase space region generated by MUPOs forming the border between the regular and chaotic region (external stickiness) and also the bouncing-ball region (internal stickiness).}
    \label{mixed_state_pic}
\end{figure}

\begin{figure*}
    \includegraphics[width=\linewidth,height=2.5in]{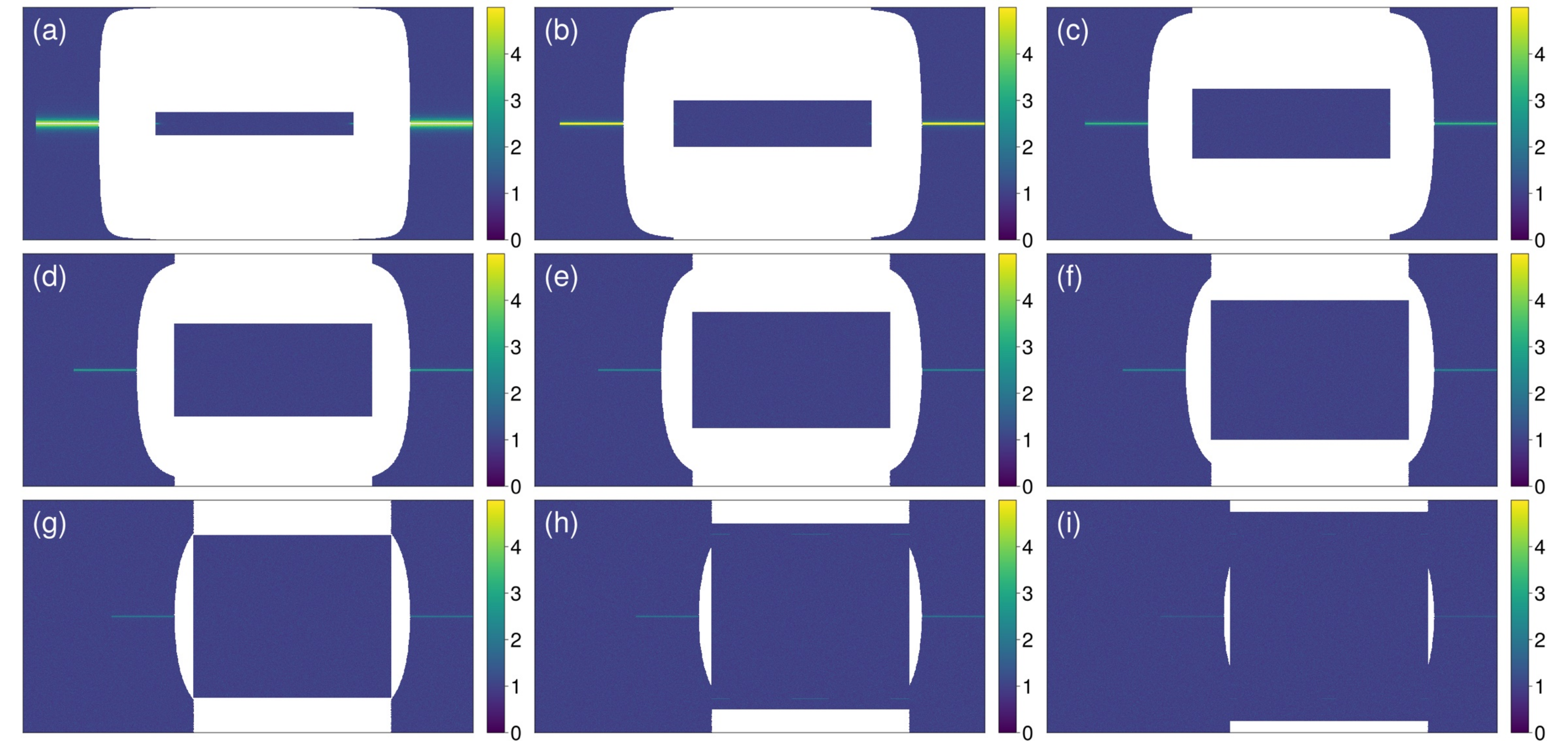}
    \caption{S plots \eqref{S_eq} for $w=0.1-0.9$, labeled $(a)-(i)$, showing stickiness around bouncing-ball modes for small $w$ and no apparent stickiness on the border between the regular and chaotic region. To enhance visualization, the color scale for $S$ is saturated at $5$ (i.e., values $S>5$ are shown as if $S=5$). Cells in the regular region are rendered white despite their value being S=0 (since $\sigma(\tau)=0$). The chaotic trajectory was iterated $N=10^8$ times.}
    \label{collision_density}
\end{figure*}
    
\section{Poincar\'e-Husimi functions}
\label{husimi}

To compare quantum and classical phase space structures, one may compute Wigner functions or, more efficiently, \textit{Poincaré–Husimi} (PH) functions \cite{TV1995,Backer2003}.  Wigner functions \cite{Wig1932} can take negative values, whereas PH functions yield a positive‐definite probability distribution on the Poincar\'e surface of section. The function itself is constructed from projecting the boundary function

\begin{equation}  \label{boudnary_function}
    u_{n}(s)=\Vec{n} \cdot  \Vec\nabla_{r} \psi_{n}(s),
\end{equation}

\noindent which is the normal derivative of the wavefunction $\psi_{n}(r(s))$ at arclength $s$ onto to a \curlyL-periodized coherent state \cite{Backer2003}

\begin{equation} \label{coherent_peak}
    c_{(q,p),k}(s) = \sum_{m \in \mathbb{Z}}\exp[ikp(s-q+m\mathop{\mathcal{L}})]
    \exp[-\frac{k}{2}(s-q+m\mathop{\mathcal{L}})^2].
\end{equation} 

\noindent The PH function is then constructed from the above objects as

\begin{equation}  \label{PH}
    H_{n}(q,p) = |\oint c_{(q,p),k_{n}}u_{n}(s)ds|^2.
\end{equation}

\noindent Fig.\ref{representative_states} shows examples of PH functions for a regular and chaotic state, showing that for regular states the PH function lives on an invariant torus while the chaotic state lives in the chaotic phase space region where it can be fully extended or localized. As will be explained in detail in Sec.\ref{mixed_states} there exist also states whose PH function lives partially on the regular and chaotic regions, called \textit{mixed-type states} (for examples see \cite{LLR2021,WR2023}). A typical mixed-type state is shown in Fig.\ref{mixed_state_pic} and shows signs of localization on the boundary between the chaotic and regular region.

\section{Separation of chaotic and regular eigenstates}
\label{separation}

\begin{figure}[!h]
  \begin{subfigure}[t]{0.5\linewidth}
    \begin{overpic}[width=\linewidth,height=1.2in]{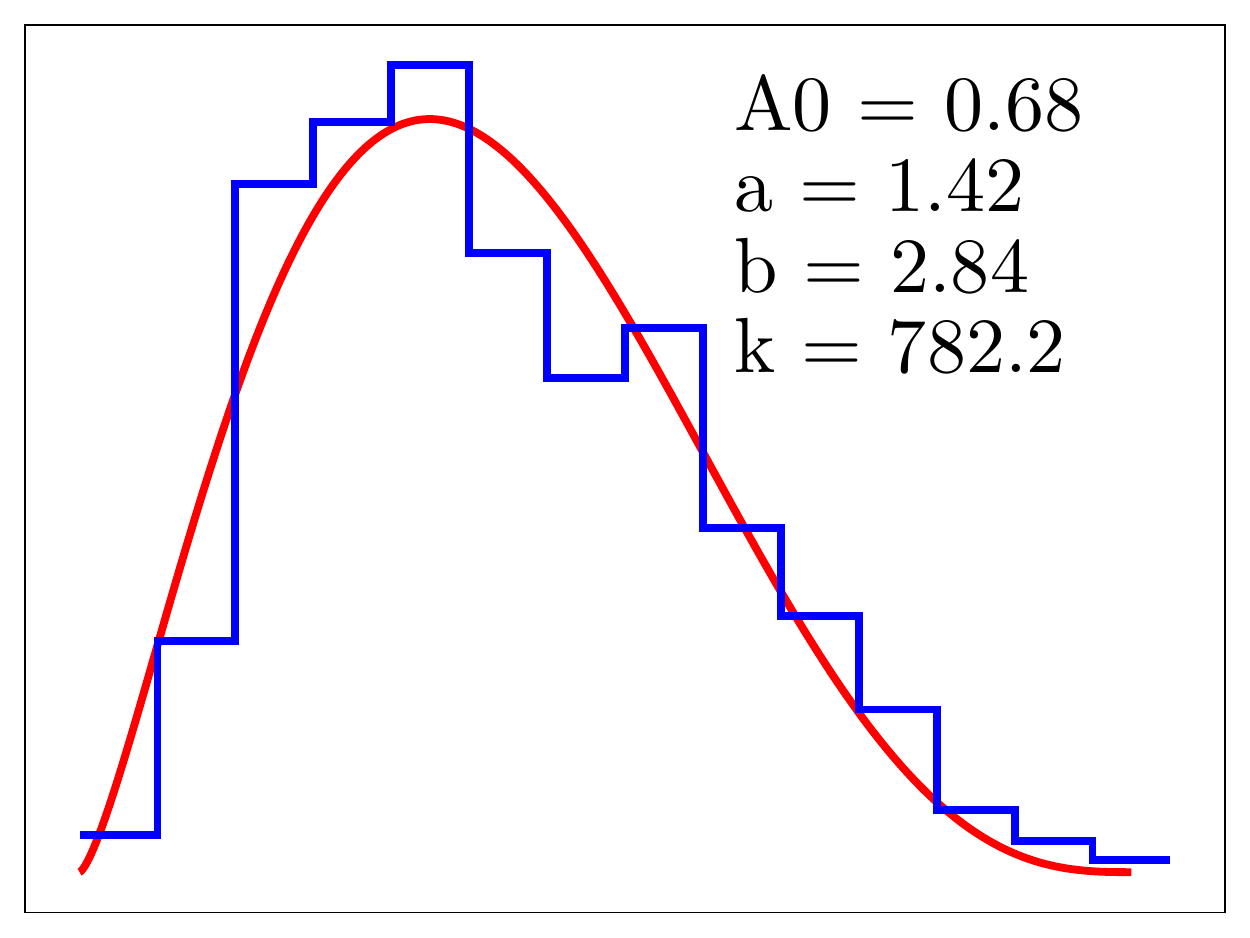}
      \put(3,57){\large (a)}
    \end{overpic}
  \end{subfigure}\hfill
  \begin{subfigure}[t]{0.5\linewidth}
    \begin{overpic}[width=\linewidth,height=1.2in]{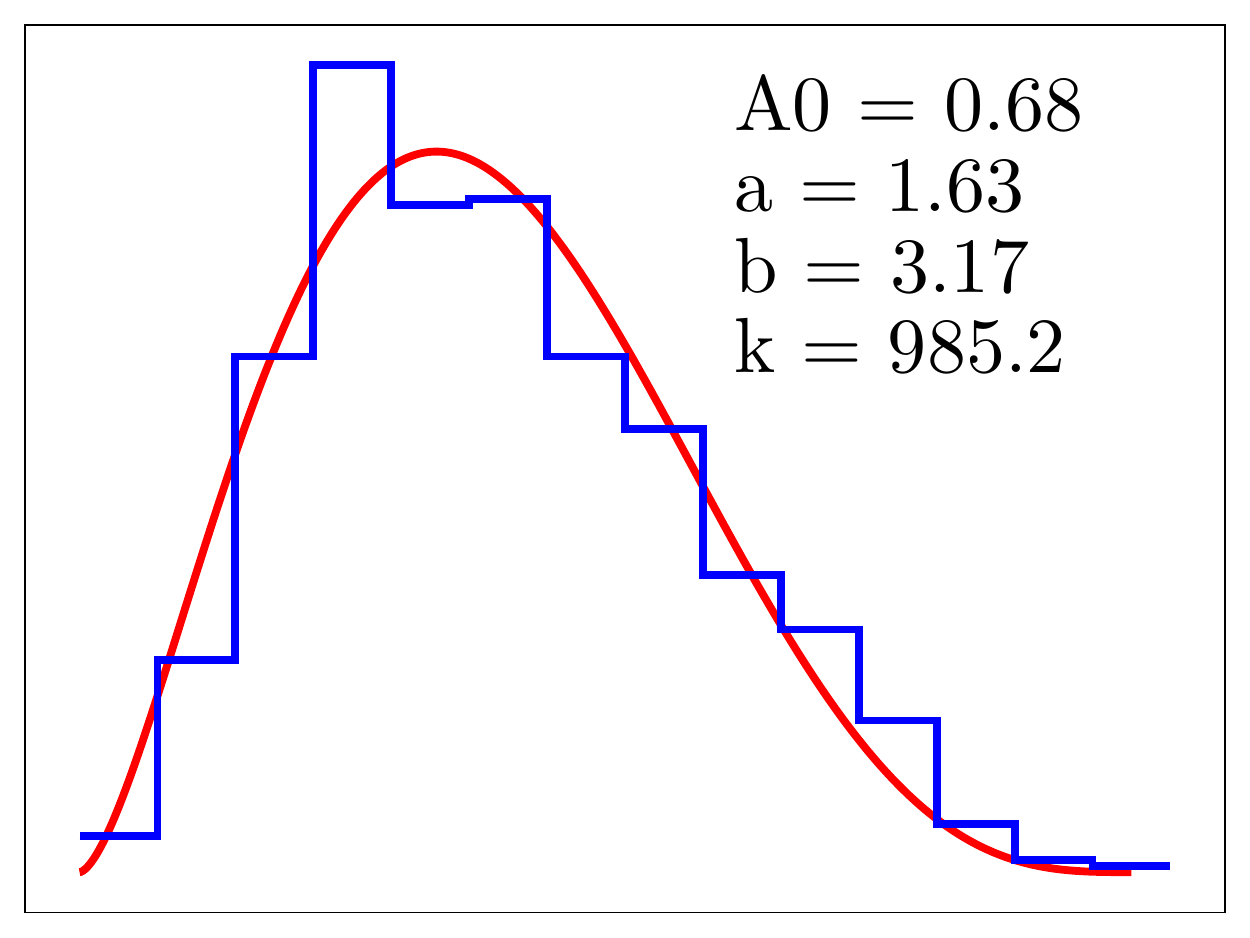}
      \put(0,57){\large\ (b)}
    \end{overpic}
  \end{subfigure}

  \begin{subfigure}[t]{0.5\linewidth}
    \begin{overpic}[width=\linewidth,height=1.2in]{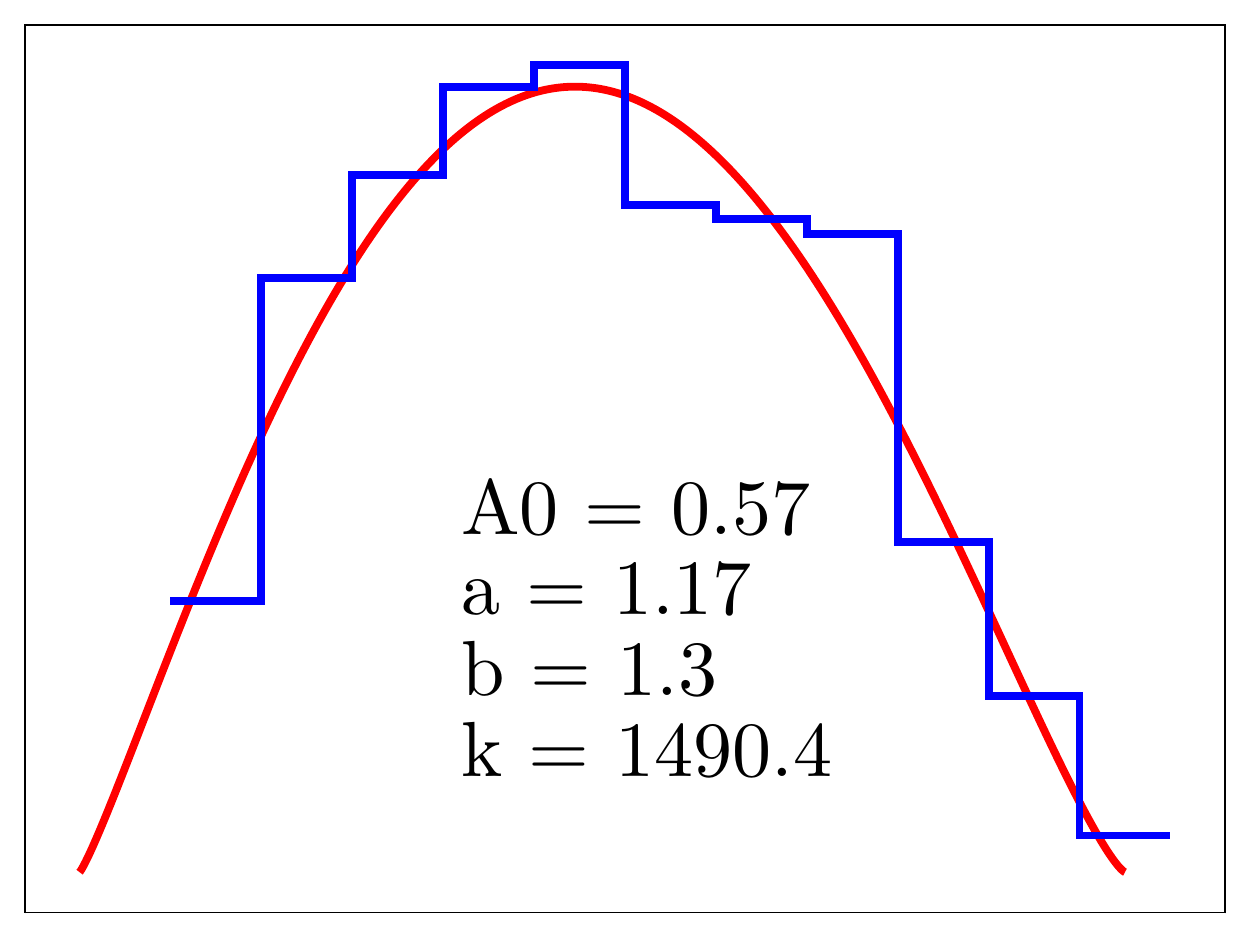}
      \put(3,55){\large (c)}
    \end{overpic}
  \end{subfigure}\hfill
  \begin{subfigure}[t]{0.5\linewidth}
    \begin{overpic}[width=\linewidth,height=1.2in]{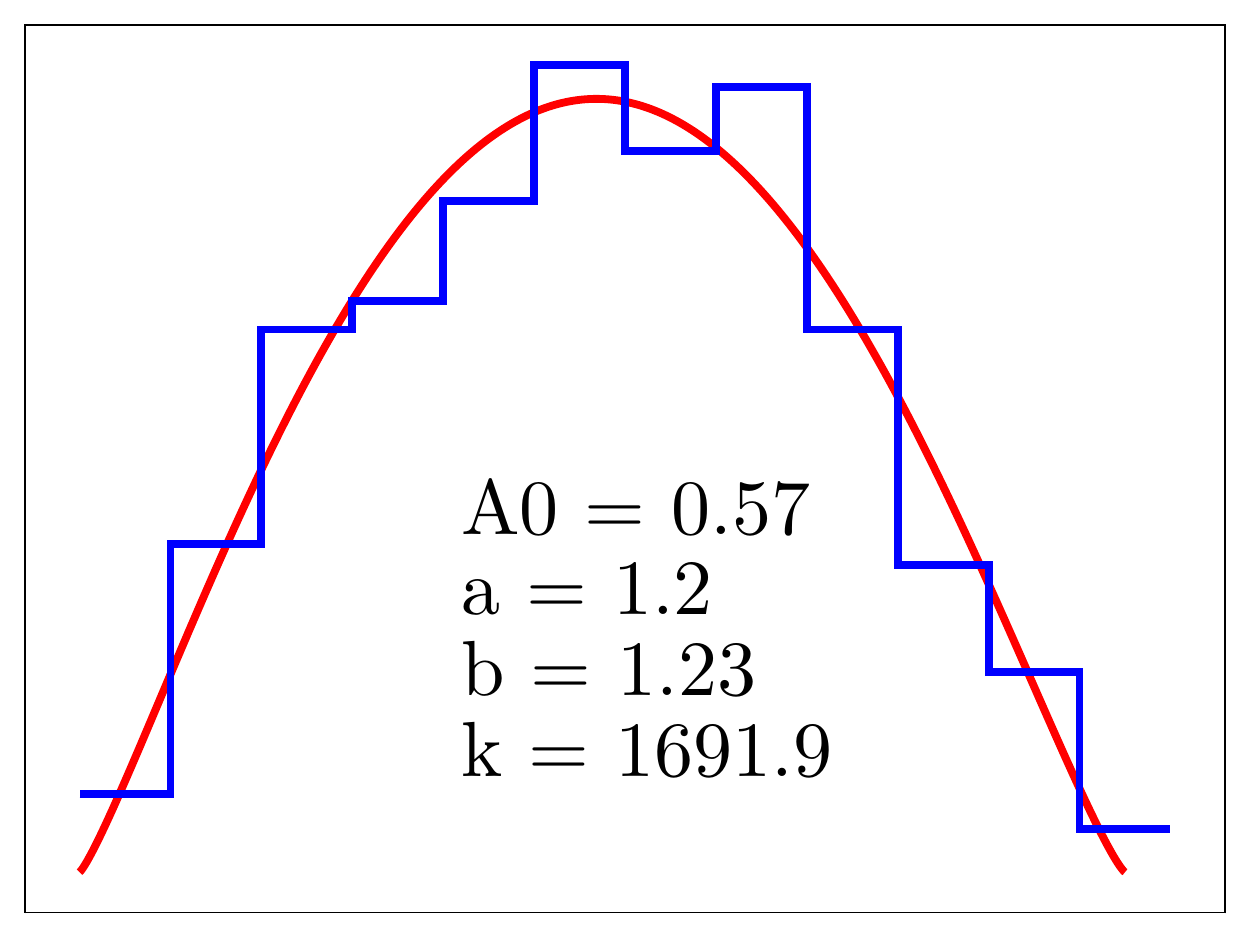}
      \put(3,55){\large (d)}
    \end{overpic}
  \end{subfigure}

  \begin{subfigure}[t]{0.5\linewidth}
    \begin{overpic}[width=\linewidth,height=1.45in]{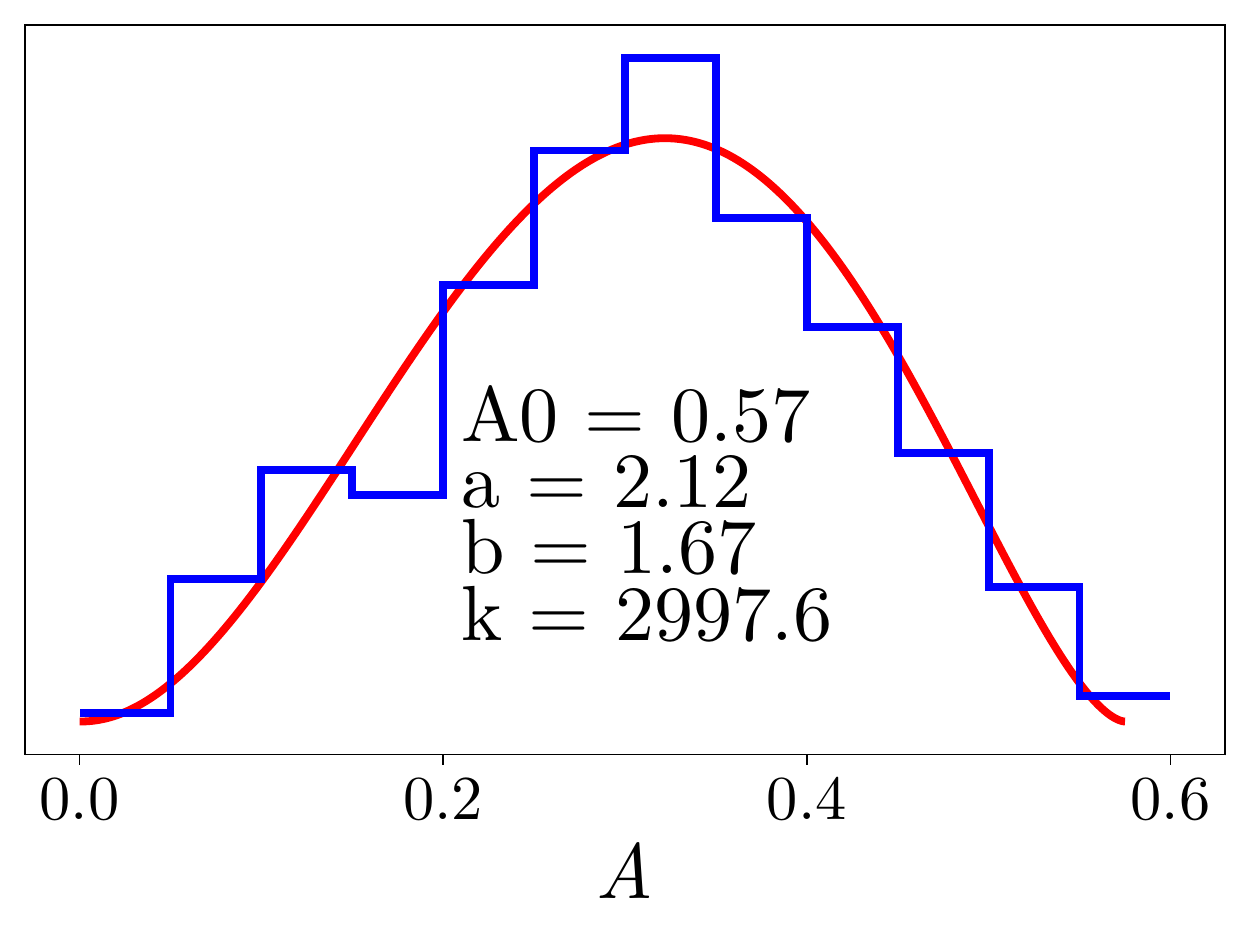}
      \put(3,70){\large (e)}
    \end{overpic}
  \end{subfigure}\hfill
  \begin{subfigure}[t]{0.5\linewidth}
    \begin{overpic}[width=\linewidth,height=1.45in]{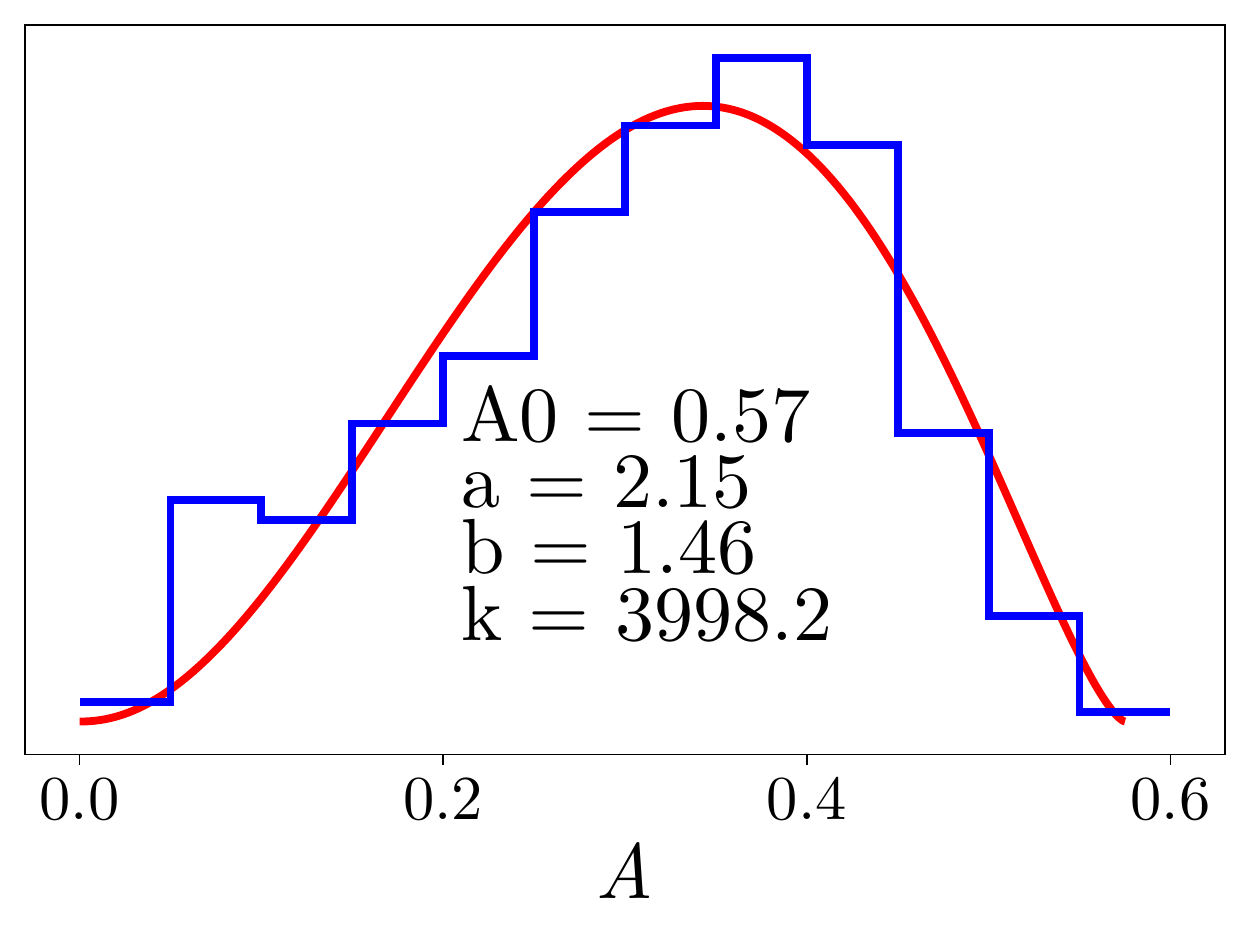}
      \put(3,70){\large (f)}
    \end{overpic}
  \end{subfigure} 

  \caption{Distribution of entropy localization measures for the mushroom billiard with stem width $w=0.1$ for chaotic states in a window centered around an increasing wavenumber $k$ containing in total $N=2000$ PH functions. (a)-(f). The fitting parameters for the beta distribution \eqref{beta_distribution} $(A_0,a,b)$ are given in each subfigure along with the center of the wavenumber window designated as $k$.}
   \label{localizations_w_0.1}
\end{figure}

To quantify localization, we must first distinguish regular from chaotic eigenstates.  We adopt the \emph{classical separation criterion} of Batistić and Robnik \cite{BatRob2013B,BatManRob2013}, which enforces that the fraction of quantum states classified as chaotic equals the classical Liouville measure of the chaotic region.  Concretely, we define the \(M\) index of the \(n\)th Poincaré–Husimi function by

\begin{figure}[!h]
  \begin{subfigure}[t]{0.5\linewidth}
    \begin{overpic}[width=\linewidth,height=1.2in]{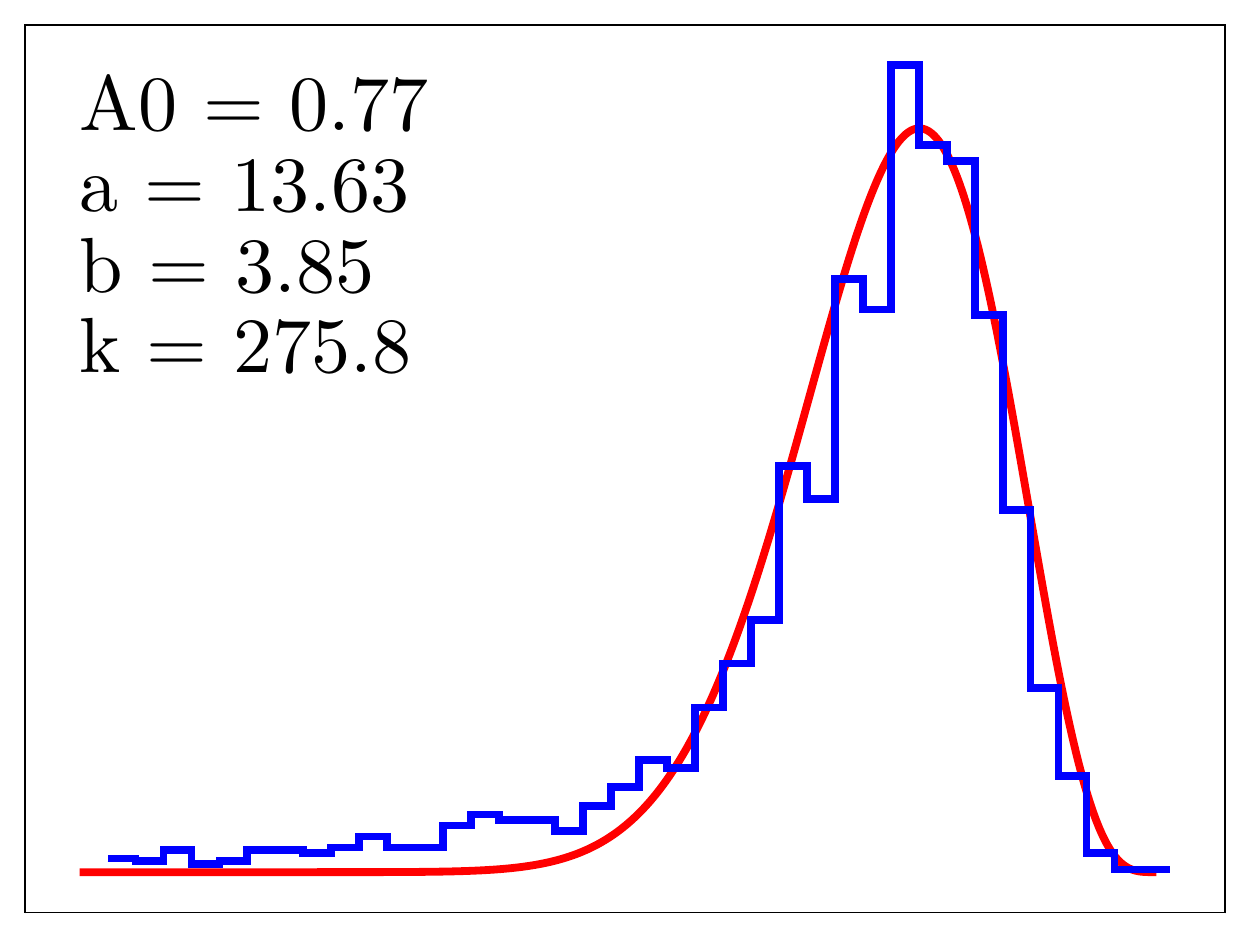}
      \put(3,30){\large (a)}
    \end{overpic}
  \end{subfigure}\hfill
  \begin{subfigure}[t]{0.5\linewidth}
    \begin{overpic}[width=\linewidth,height=1.2in]{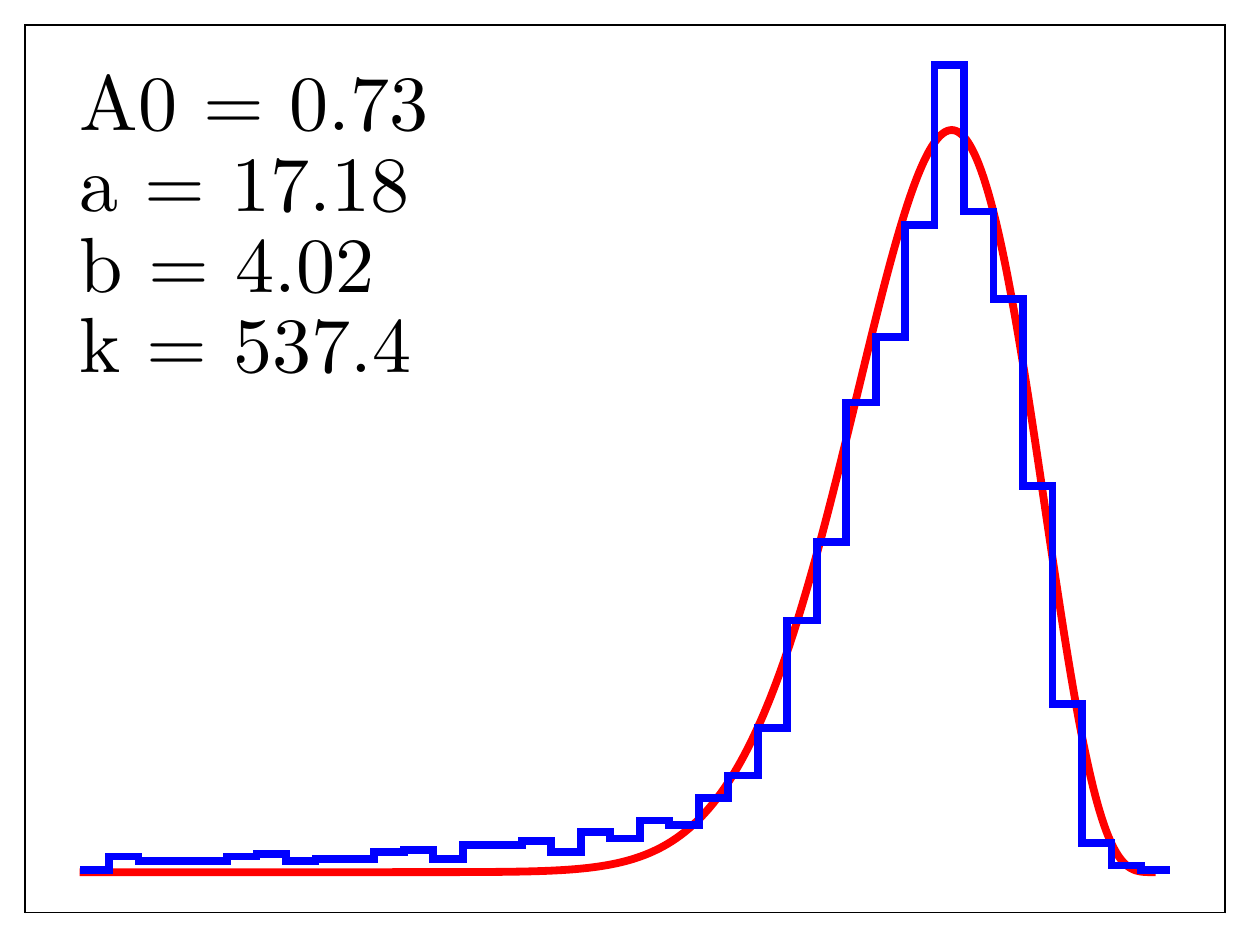}
      \put(3,30){\large\ (b)}
    \end{overpic}
  \end{subfigure}

  \begin{subfigure}[t]{0.5\linewidth}
    \begin{overpic}[width=\linewidth,height=1.2in]{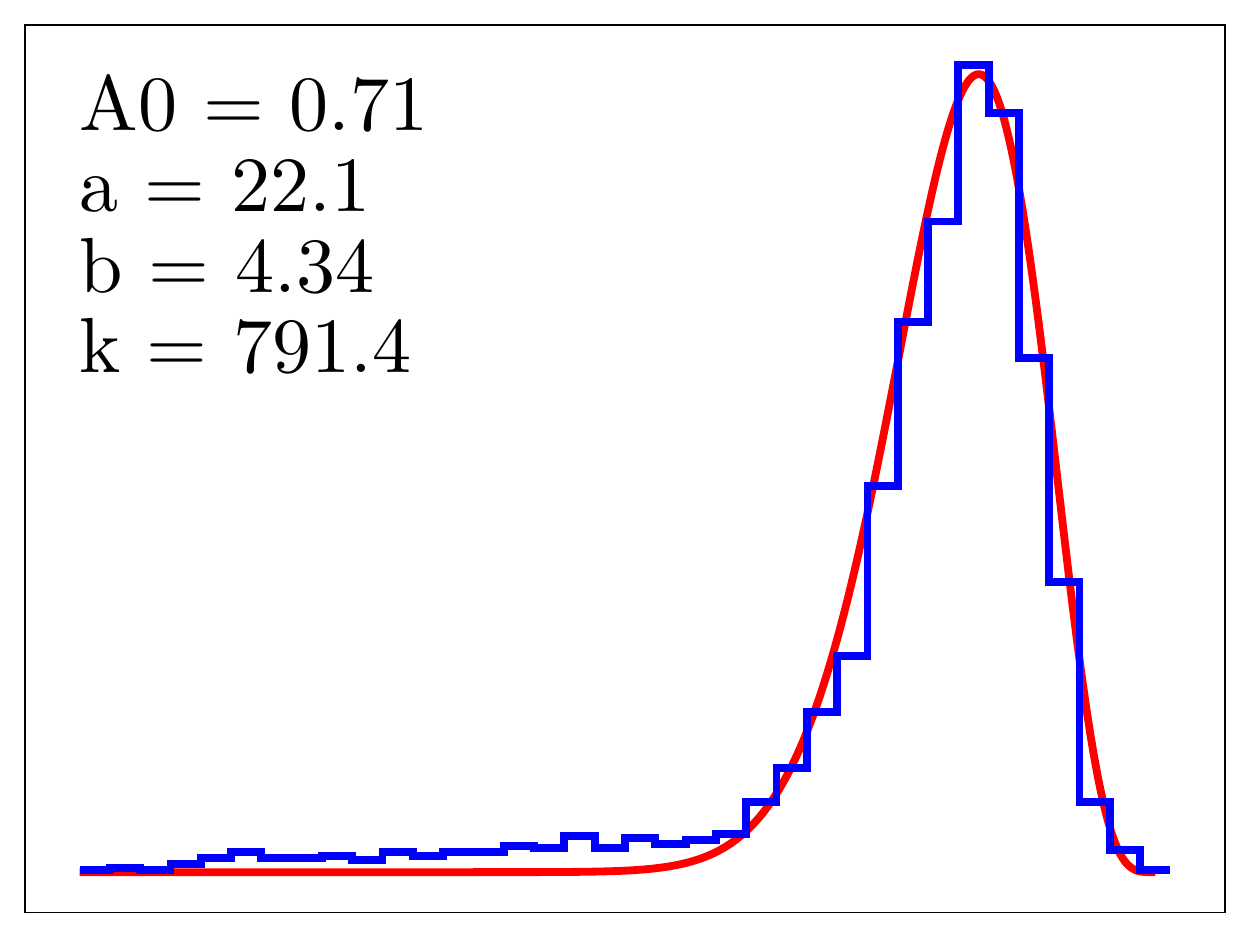}
      \put(3,30){\large (c)}
    \end{overpic}
  \end{subfigure}\hfill
  \begin{subfigure}[t]{0.5\linewidth}
    \begin{overpic}[width=\linewidth,height=1.2in]{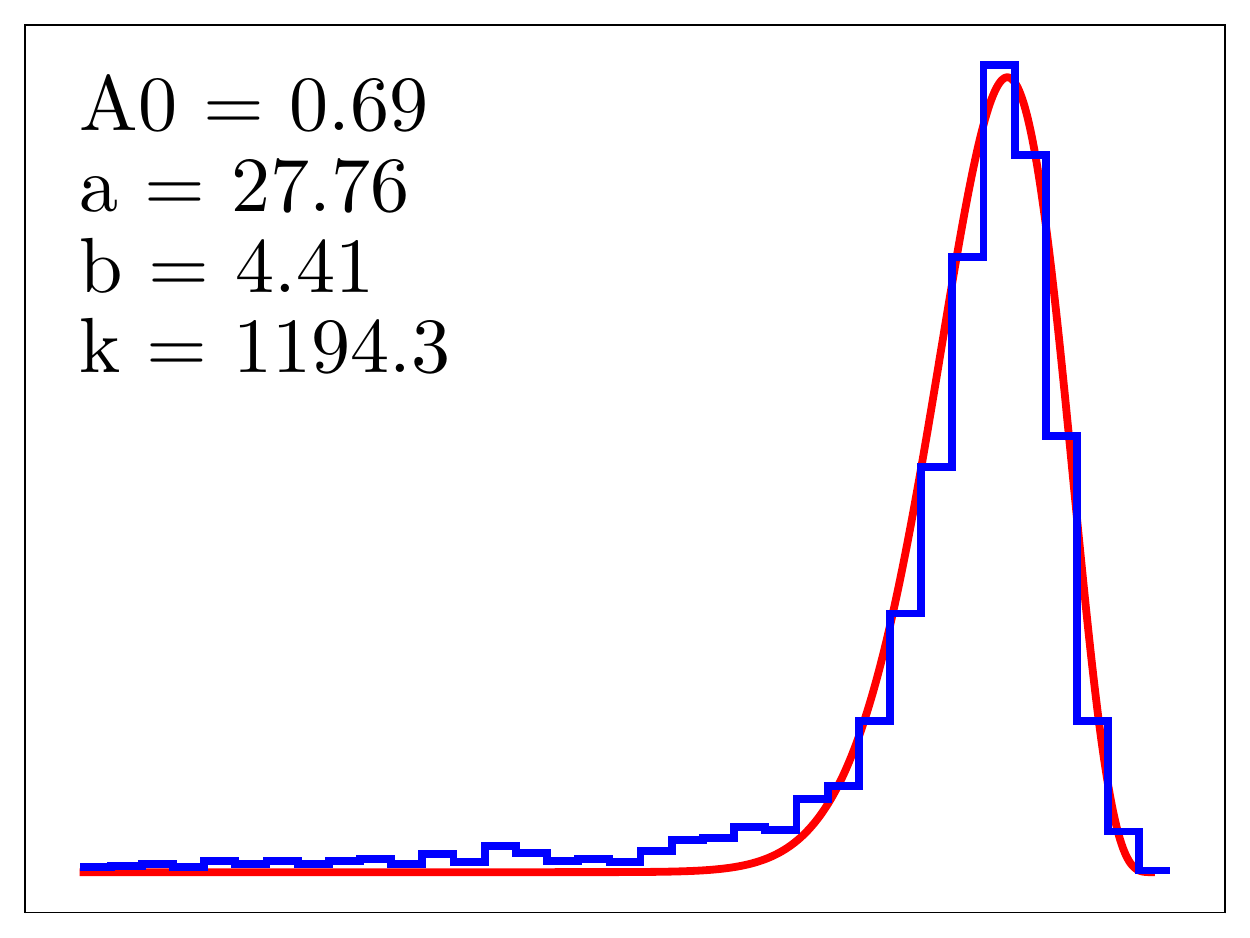}
      \put(3,30){\large (d)}
    \end{overpic}
  \end{subfigure}

  \begin{subfigure}[t]{0.5\linewidth}
    \begin{overpic}[width=\linewidth,height=1.45in]{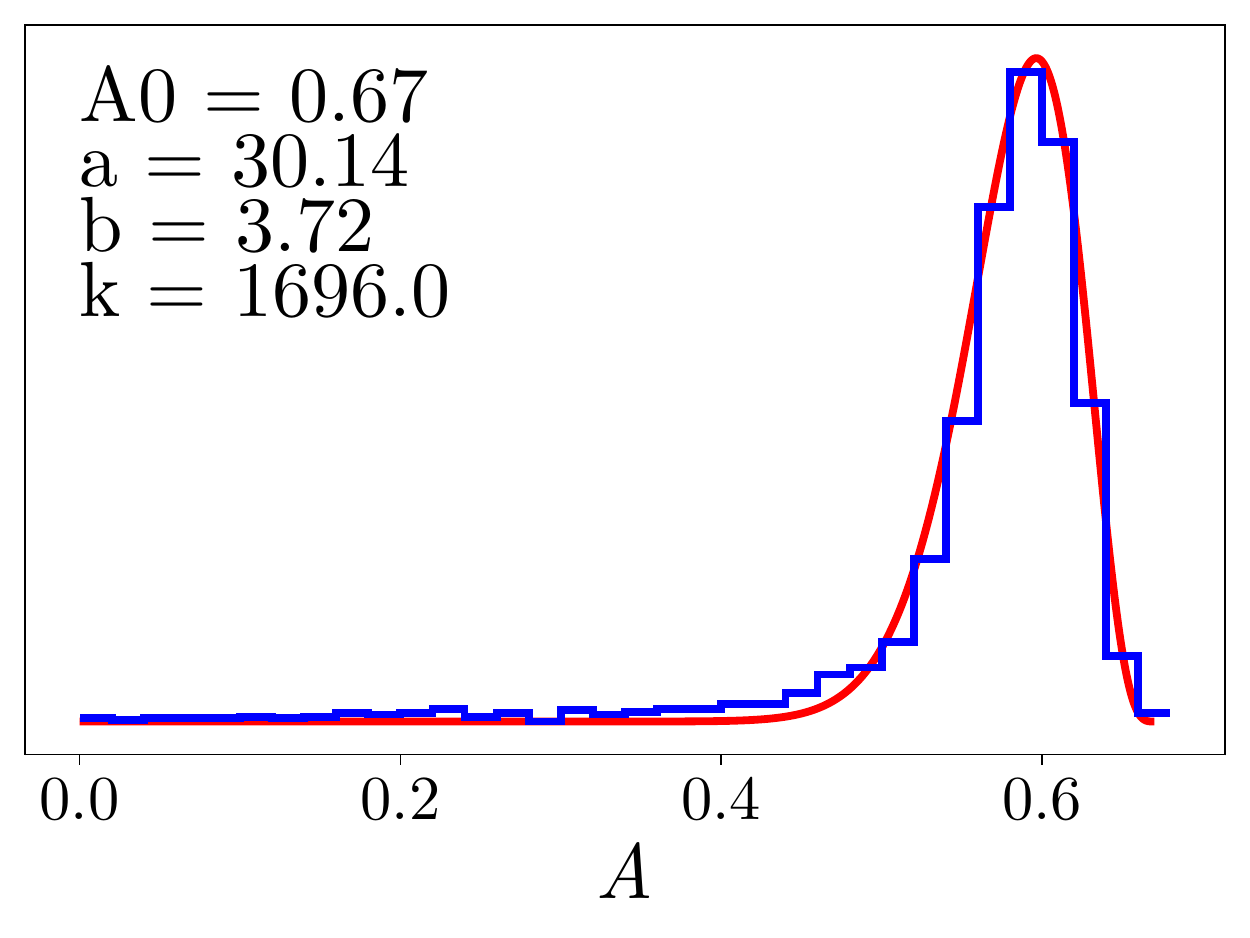}
      \put(3,45){\large (e)}
    \end{overpic}
  \end{subfigure}\hfill
  \begin{subfigure}[t]{0.5\linewidth}
    \begin{overpic}[width=\linewidth,height=1.45in]{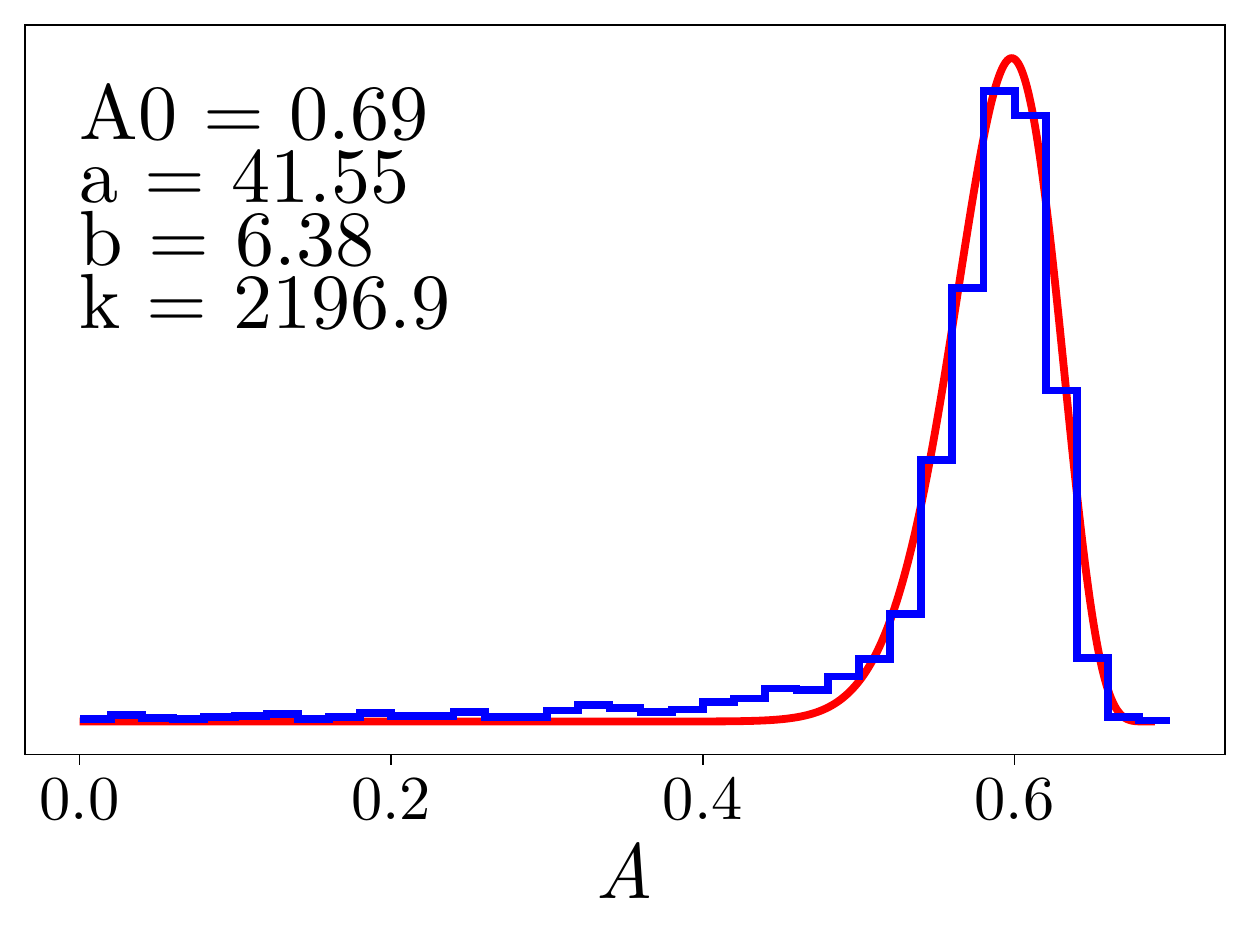}
      \put(4,45){\large (f)}
    \end{overpic}
  \end{subfigure} 

  \caption{Distribution of entropy localization measures for the mushroom billiard with stem width $w=0.6$ for chaotic states in a window centered around an increasing wavenumber $k$ containing in total $N=2000$ PH functions. (a)-(f). The fitting parameters for the beta distribution \eqref{beta_distribution} $(A_0,a,b)$ are given in each subfigure along with the center of the wavenumber window designated as $k$.}
  \label{localizations_w_0.6}
\end{figure}

\begin{equation}\label{m_index}
  M_n \;=\; \sum_{i,j} C_{ij}\,H_n\bigl(q_i,p_j\bigr)\,,
\end{equation}

\noindent where \(H_n(q_i,p_j)\) is the discrete PH value in the \((i,j)\)th cell of the \(N_q\times N_p\) grid, and
\[
  C_{ij} = 
    \begin{cases}
      +1, & \text{if cell \((i,j)\) is visited by a chaotic trajectory},\\
      -1, & \text{otherwise}.
    \end{cases}
\]
The classical map was sampled by iterating a single chaotic orbit for \(10^8\) collisions. 

An eigenstate \(n\) is then declared chaotic if \(M_n \ge M_{\rm th}\), and regular otherwise.  The threshold \(M_{\rm th}\) is chosen so as to match the quantum fraction of chaotic levels to the known analytic value \(\mu_{\rm ch}\) of the chaotic phase‐space volume \cite{BarBet2007,OreLozRobYan2025}.  Equivalently, \(M_{\rm th}\) minimizes
\[
  f(M_{\rm th})
  \;=\;
  \biggl|\frac{N_{\rm ch}(M_{\rm th})}{N} \;-\; \mu_{\rm ch}\biggr|,
\]
where \(N_{\rm ch}(M_{\rm th})\) is the number of states with \(M_n\ge M_{\rm th}\) among the \(N\) total levels. For this minimization we decreased $M_{th}$ until the relative error was reached as

\begin{figure}[ht]
  \begin{subfigure}[t]{0.5\linewidth}
    \begin{overpic}[width=\linewidth,height=1.2in]{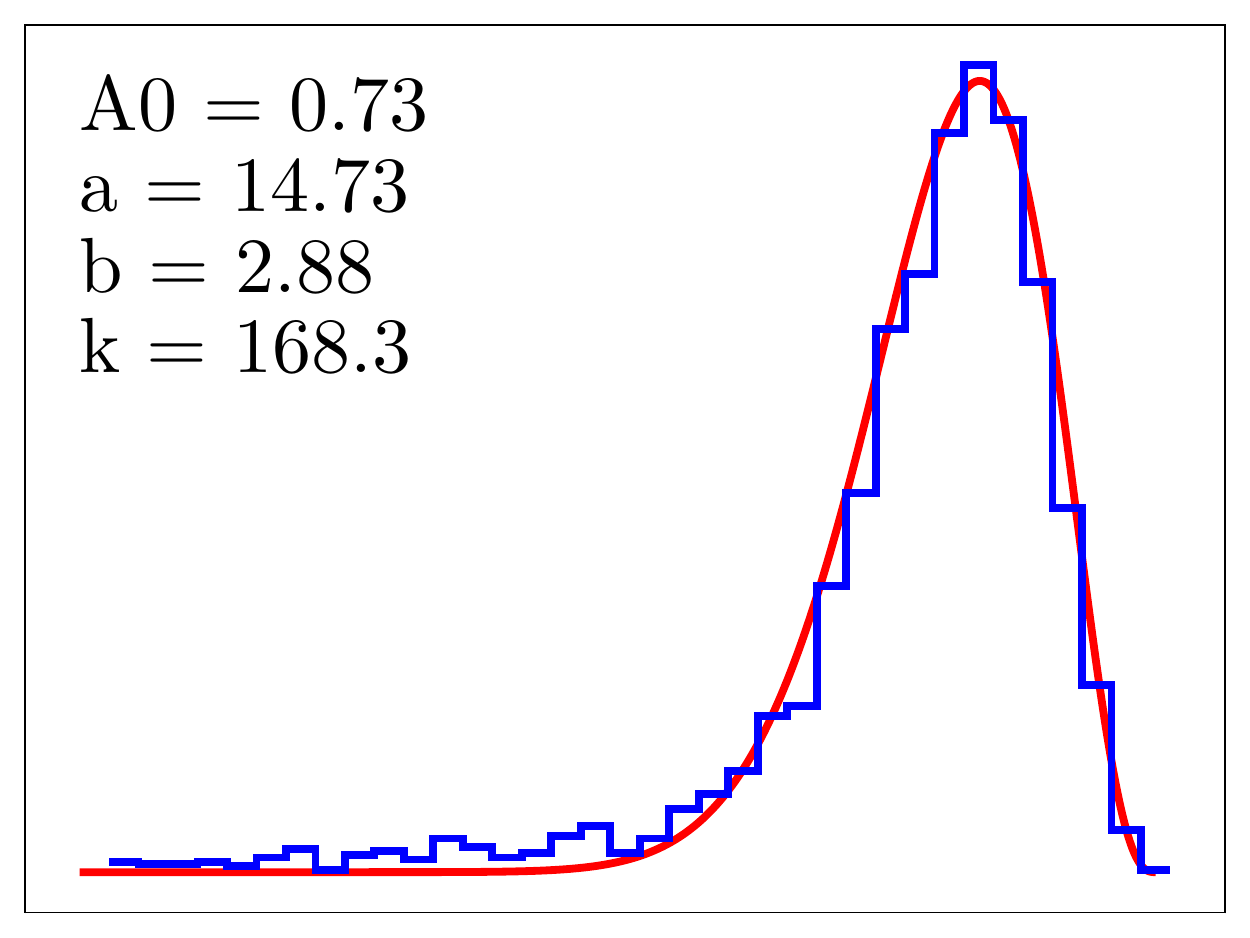}
      \put(3,30){\large (a)}
    \end{overpic}
  \end{subfigure}\hfill
  \begin{subfigure}[t]{0.5\linewidth}
    \begin{overpic}[width=\linewidth,height=1.2in]{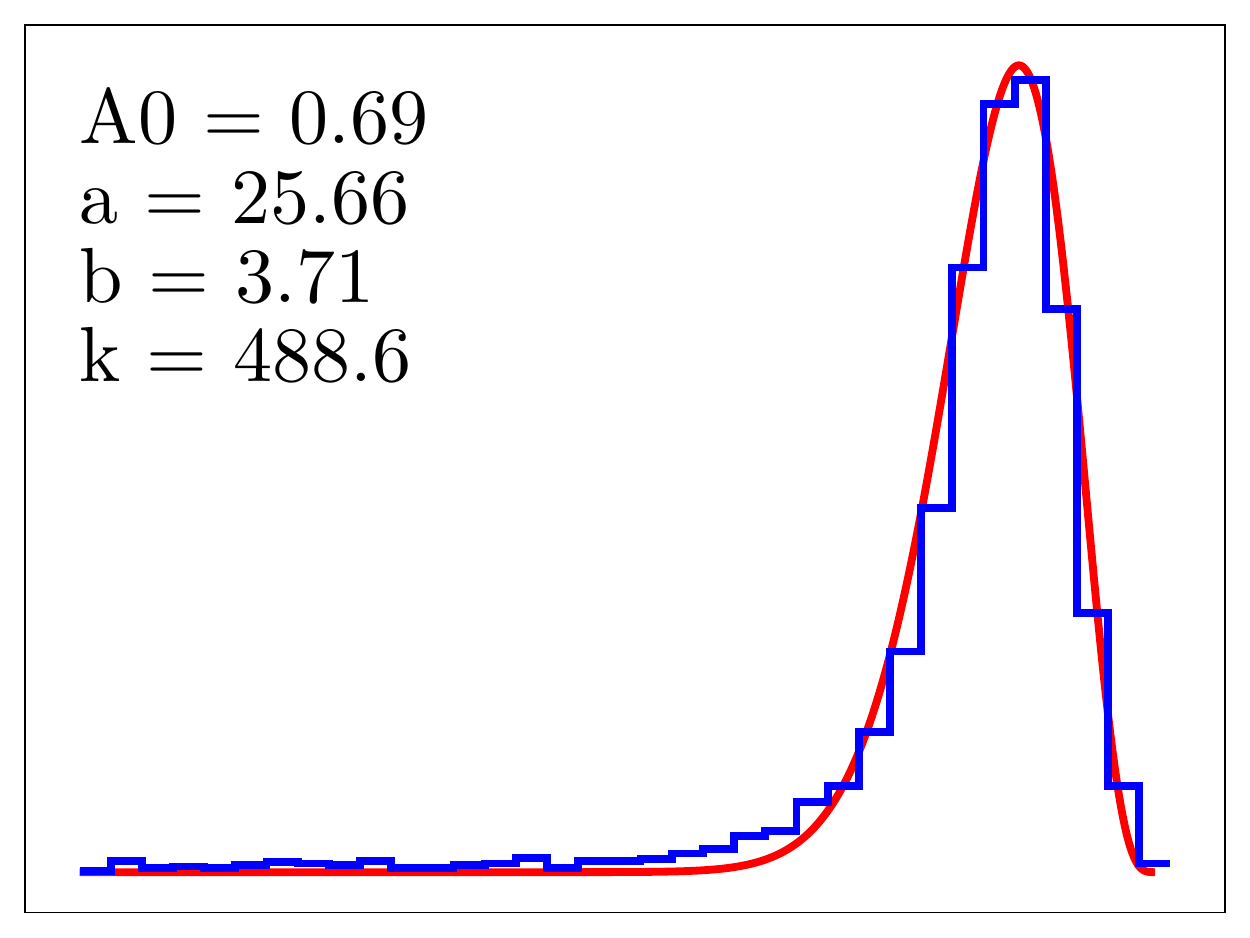}
      \put(3,30){\large\ (b)}
    \end{overpic}
  \end{subfigure}

  \begin{subfigure}[t]{0.5\linewidth}
    \begin{overpic}[width=\linewidth,height=1.2in]{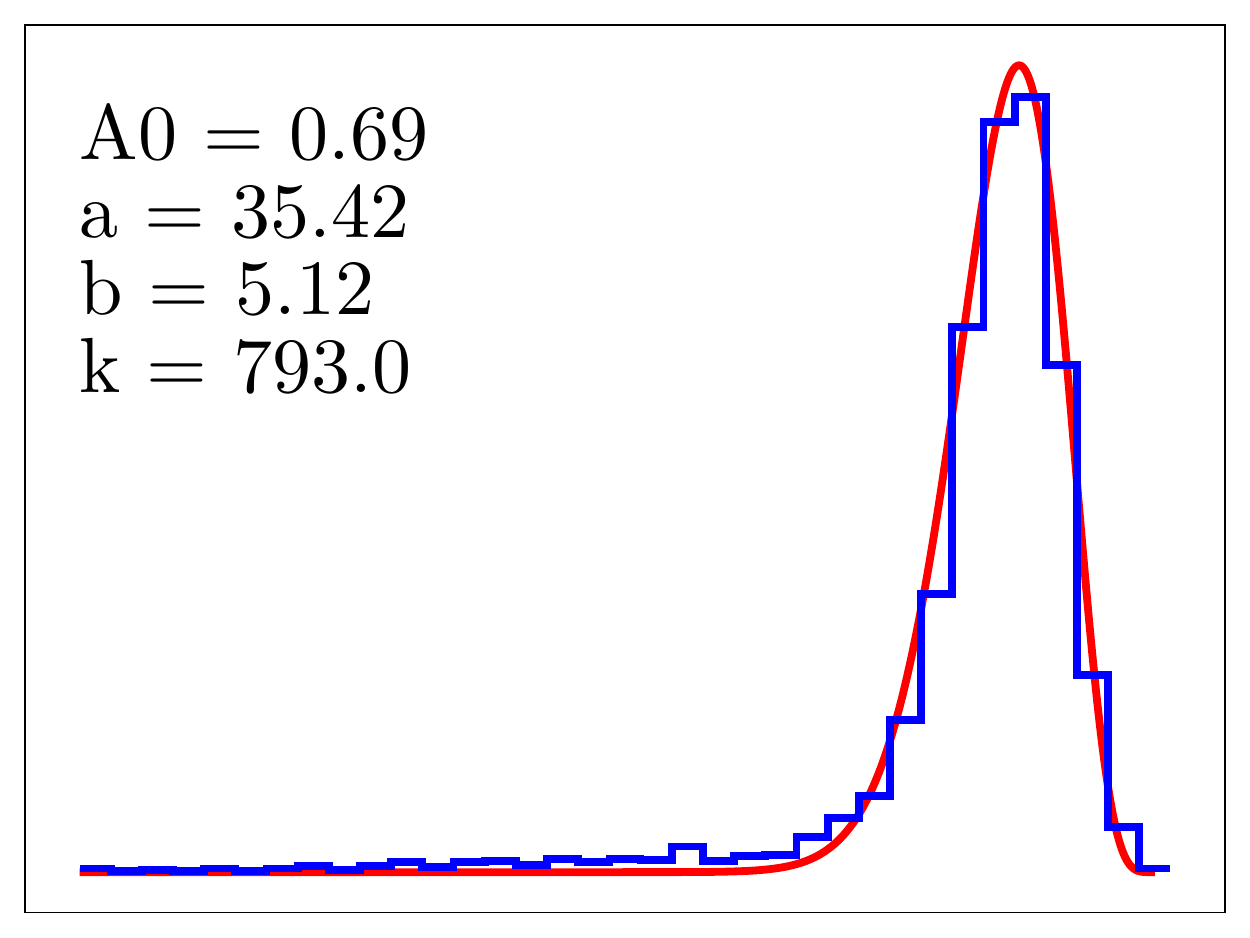}
      \put(3,30){\large (c)}
    \end{overpic}
  \end{subfigure}\hfill
  \begin{subfigure}[t]{0.5\linewidth}
    \begin{overpic}[width=\linewidth,height=1.2in]{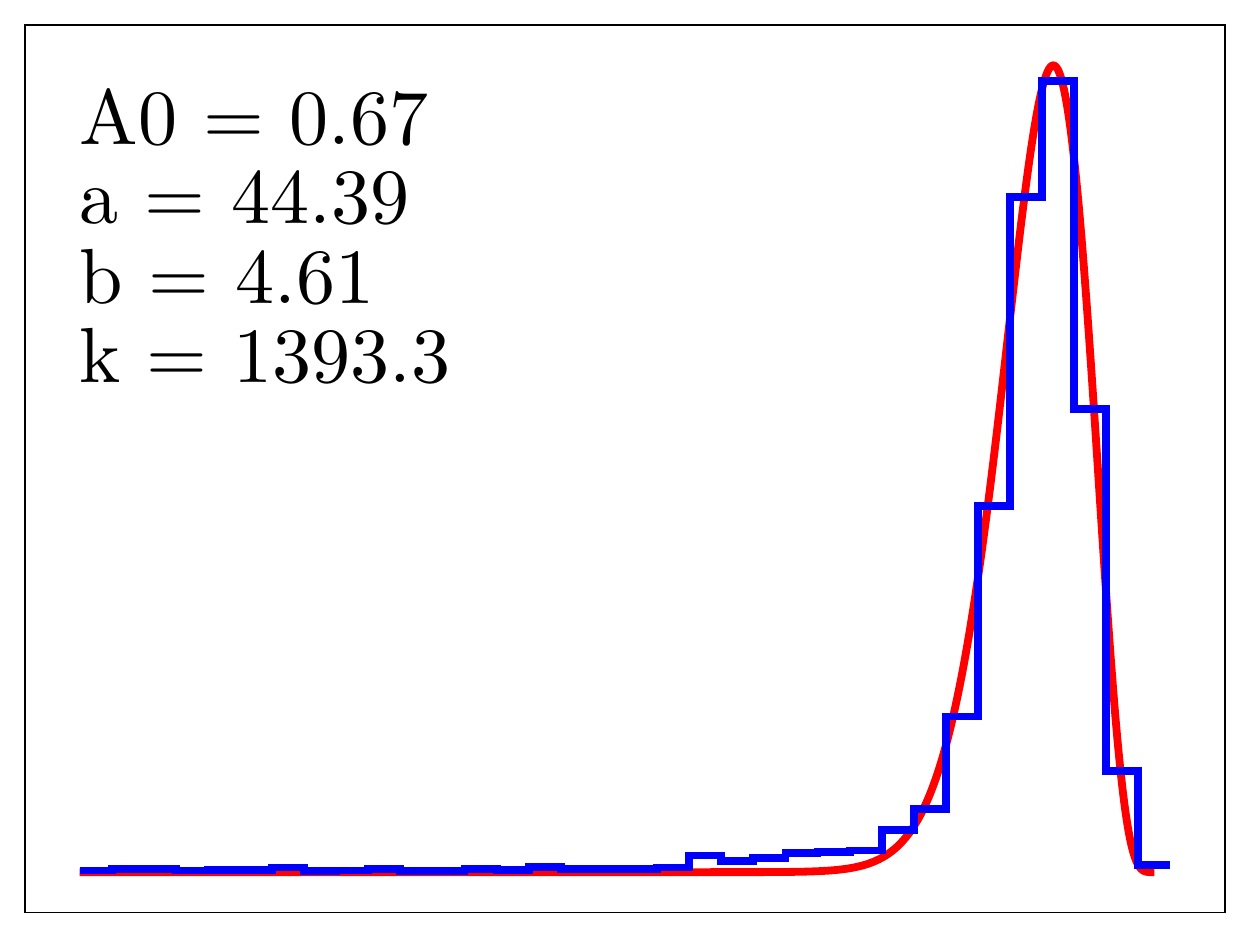}
      \put(3,30){\large (d)}
    \end{overpic}
  \end{subfigure}

  \begin{subfigure}[t]{0.5\linewidth}
    \begin{overpic}[width=\linewidth,height=1.45in]{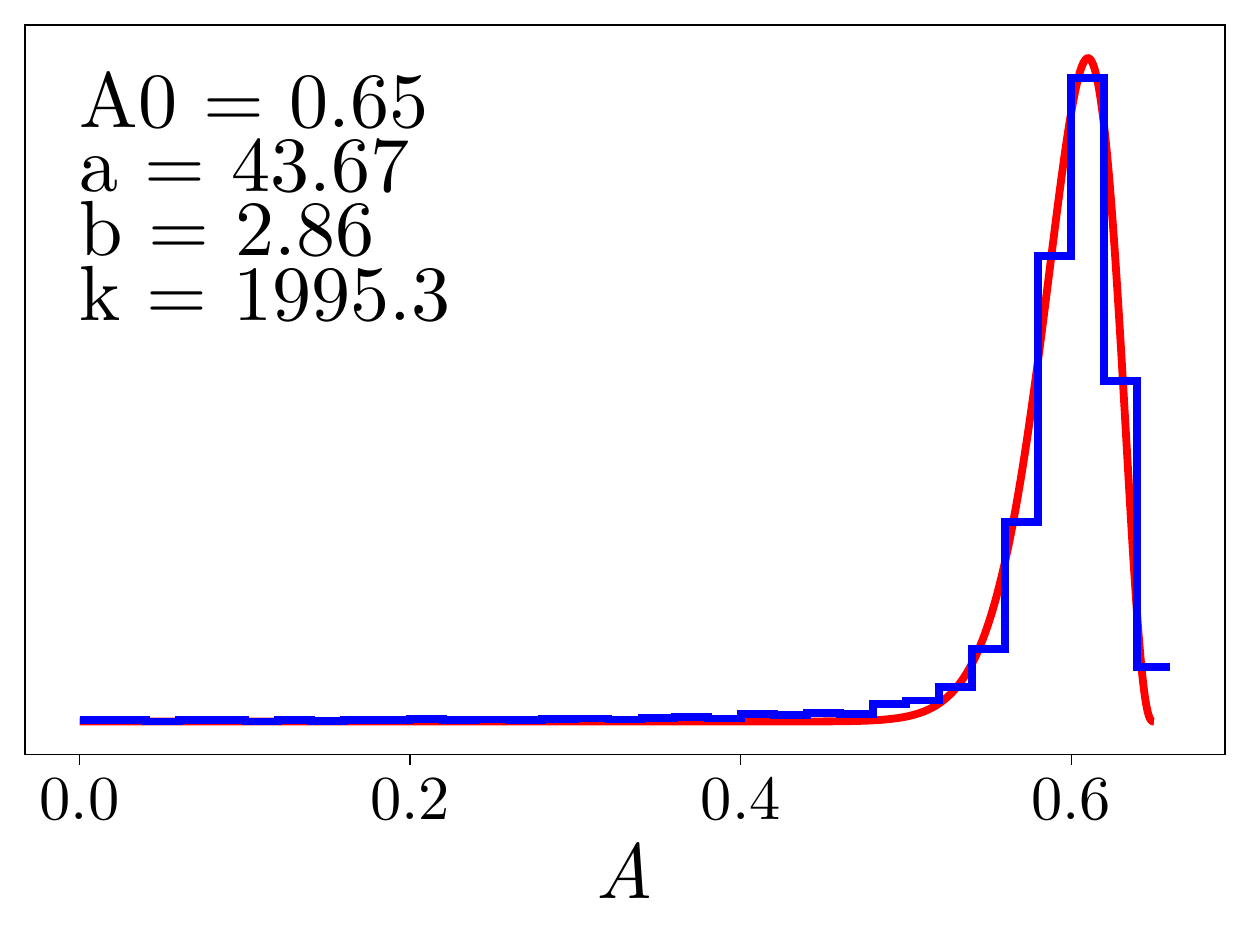}
      \put(3,45){\large (e)}
    \end{overpic}
  \end{subfigure}\hfill
  \begin{subfigure}[t]{0.5\linewidth}
    \begin{overpic}[width=\linewidth,height=1.45in]{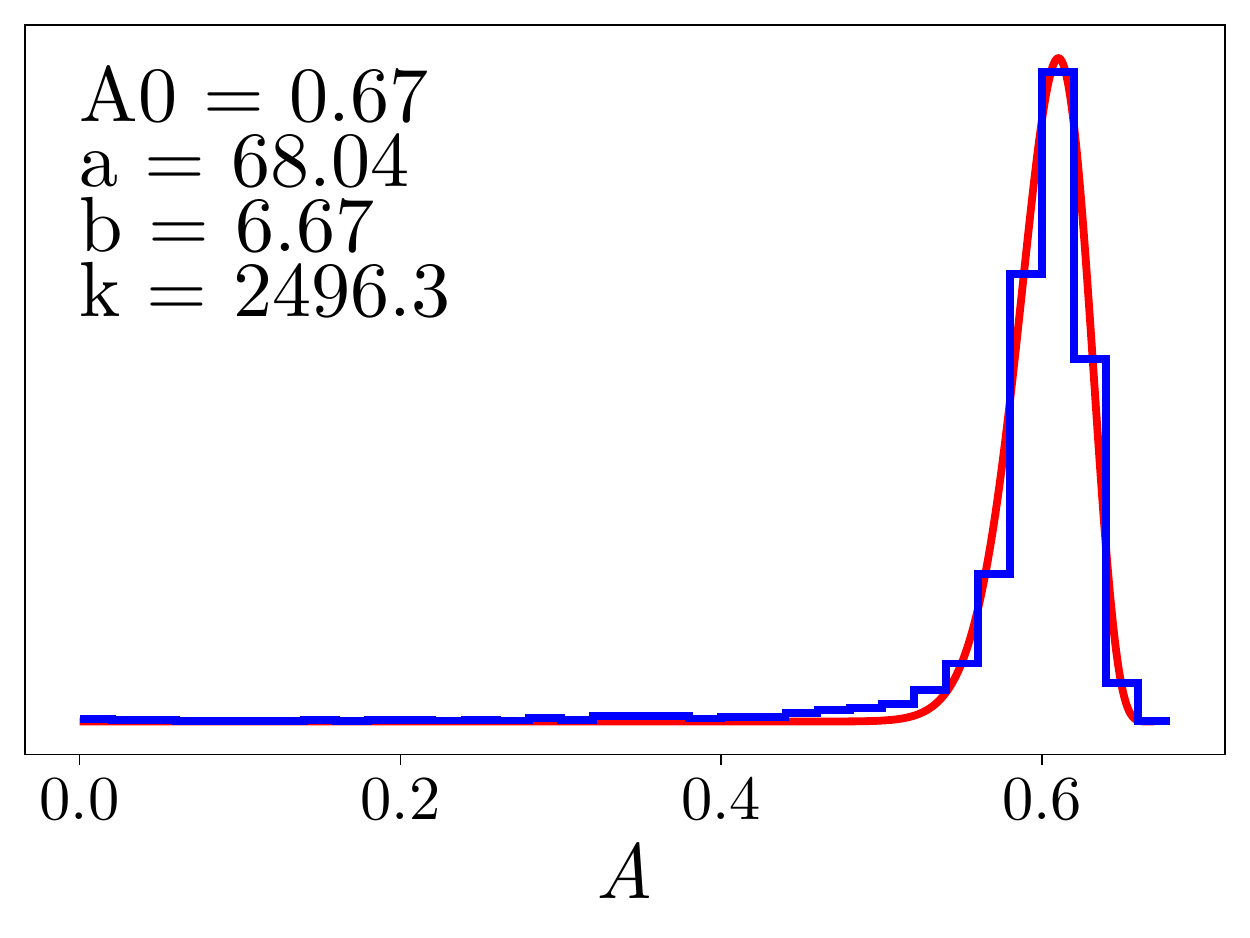}
      \put(4,45){\large (f)}
    \end{overpic}
  \end{subfigure} 

  \caption{Distribution of entropy localization measures for the mushroom billiard with stem width $w=0.9$ for chaotic states in a window centered around an increasing wavenumber $k$ containing in total $N=2000$ PH functions. (a)-(f). The fitting parameters for the beta distribution \eqref{beta_distribution} $(A_0,a,b)$ are given in each subfigure along with the center of the wavenumber window designated as $k$.}
  \label{localizations_w_0.9}
\end{figure}

\begin{equation} \label{numerical_separation_criterion}
    \biggl| \frac{\frac{N_{\rm ch}(M_{\rm th})}{N}-\mu_{ch}}{\mu_{ch}}\biggl|<0.03.
\end{equation}

We also searched for diagnostic indicators of stickiness using recurrence time statistics on the Poincar\'e surface of section, visualized with S-plots \cite{Lozej2020}. For each cell $(i,j)$ in the grid we calculated:

\begin{figure*}[]
    \includegraphics[width=\linewidth]{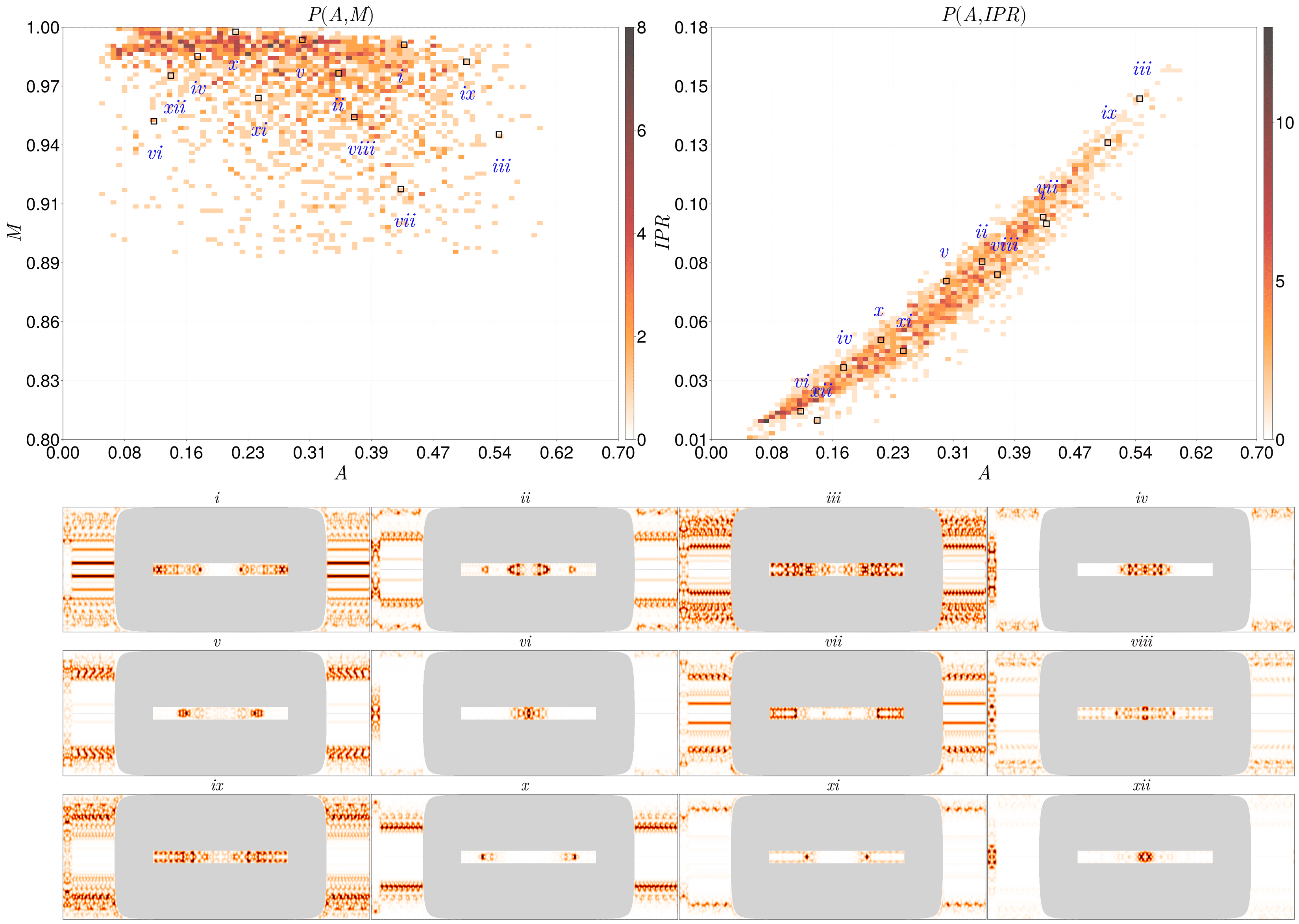}
    \put(-0.92\linewidth,0.4\linewidth){\textbf{(a)}}
    \put(-0.1\linewidth,0.4\linewidth){\textbf{(b)}}
    \caption{Joint probability density $P(A,M)$ (a) and $P(A,IPR)$ (b) for chaotic states for stem width $w=0.1$ in the wavenumber window centered at $k \approx 2000.0$ containing approximately $10000$ PH functions in total (here showing only chaotic). (Below) Some representatives from the PH function ensemble showing pronounced effects of dynamical localization for most of the chaotic states. The grey area in PH function plots is the regular component.The scale in both heatmaps is linear, signifying counts of PH functions in a \([A,A+\delta A]\times[M,M+\delta M]\) or \([A,A+\delta A]\times[\mathrm{IPR},\mathrm{IPR}+\delta\mathrm{IPR}]\) bin, respectively. PH function $(xii)$ shows complete localization of a chaotic eigenstate on a periodic orbit of length 2, namely the bounces between the circular cap top and the bottom in the stem.}
    \label{combined_w_0.1}
\end{figure*}

\begin{equation} \label{S_eq}
    S(i,j) = \frac{\sigma(i,j)}{\tau(i,j)},
\end{equation}

\noindent where $\sigma(i,j)$ and $\tau(i,j)$ are the standard deviation of recurrence times and the mean recurrence time for the cell indexed by $(i,j)$. Fig.\ref{collision_density} shows the values of $S$ in the chaotic region for all studied stem widths $w$, showing that outside the small sticky area around the bouncing-ball modes we have $S=1$, giving an exponential distribution of recurrence times. Values of $0<S<1$ can theoretically occur in this region and do not indicate regularity.

\section{Localization measures}
\label{localizations}

\subsection{Definitions}

There are many possible measures of phase‐space localization.  Here we focus on two widely used quantities: the entropy localization measure \(A_n\) and the inverse participation ratio \(\mathrm{IPR}_n\).  Let \(H_n(q_i,p_j)\) denote the \(n\)th Poincaré–Husimi function sampled on a grid of \(N\) cells, and assume it is normalized so that \(\sum_{i,j}H_n(q_i,p_j)=1\).  Further let \(N_c\) be the number of cells belonging to the classical chaotic region.  We then define

\begin{equation} \label{localization_entropy}
  A_n \;=\;\frac{1}{N_c}
    \exp\!\biggl(
      -\sum_{i,j}H_n(q_i,p_j)\,\ln H_n(q_i,p_j)
    \biggr)\,,
\end{equation}

\begin{figure*}[ht]
    \includegraphics[width=\linewidth]{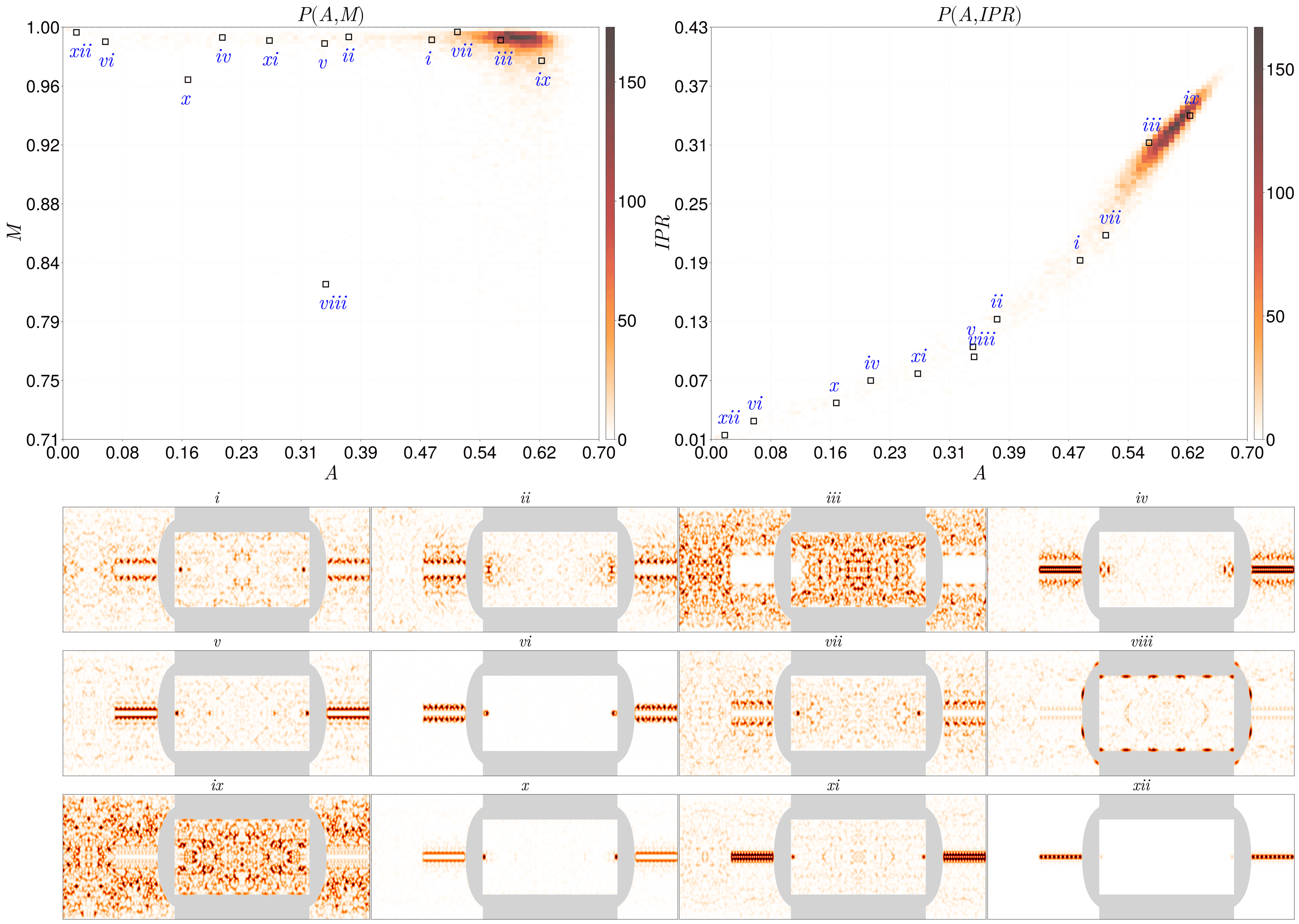}
    \put(-0.92\linewidth,0.4\linewidth){\textbf{(a)}}
    \put(-0.1\linewidth,0.4\linewidth){\textbf{(b)}}
    \caption{Joint probability density $P(A,M)$ (a) and $P(A,IPR)$ (b)  for chaotic states for stem width $w=0.6$ in the wavenumber window centered at $k \approx 2000.0$ containing approximately $10000$ PH functions in total (here showing only chaotic). (Below) Some representatives from the PH function ensemble showing pronounced effects of dynamical localization for most of the chaotic states. The gray area in PH function plots is the regular component. The scale in both heatmaps is linear, signifying counts of PH functions in a \([A,A+\delta A]\times[M,M+\delta M]\) or \([A,A+\delta A]\times[\mathrm{IPR},\mathrm{IPR}+\delta\mathrm{IPR}]\) bin, respectively. PH function (viii) shows an example of a mixed-type state showing increased density in the MUPO that generates internal stickiness.}
    \label{combined_w_0.6}
\end{figure*}

\noindent which quantifies the effective volume occupied by the PH density within the chaotic domain.  In the extreme case of a fully extended state, \(H_n=1/N_c\) on the chaotic cells and zero elsewhere, so that \(-\sum H_n\ln H_n=\ln N_c\) and hence \(A_n=(1/N_c)\,e^{\ln N_c}=1\).  Conversely, if the state is maximally localized in a single cell, \(H_n=1\) at that cell and zero elsewhere, then \(-\sum H_n\ln H_n=0\) and \(A_n=(1/N_c)\,e^0=1/N_c\ll1\).

Likewise, the inverse participation ratio is
\begin{equation} \label{IPR}
  \mathrm{IPR}_n
  \;=\;
  \frac{1}{N_c}
  \,\frac{1}{\displaystyle
    \sum_{i,j}H_n(q_i,p_j)^{2}
  }
  \,,
\end{equation}

\noindent so that \(\mathrm{IPR}_n=1\) for a state uniformly spread over all \(N\) cells, and \(\mathrm{IPR}_n\ll1\) for strongly localized states.

\subsection{Distribution of localization measures}
\label{dist_loc_measures}

We have looked at an ensemble of about $2000$ consecutive eigenstates around certain $k$ and analyzed the statistics of entropy localization measure $A$ of chaotic eigenstates (after separation). It is found that for almost all $w$ under study we found a good agreement of the distribution of localization measures $P(A_n)$ of chaotic states with the so-called \textit{beta distribution} \cite{BLR2020}

\begin{equation} \label{beta_distribution}
    P(A)=CA^a(A_0-A)^b,
\end{equation}

\noindent where $A_0$ is the upper limit of the interval where this distribution is defined, $a$ and $b$ are positive system-dependent parameters and $C$ is determined by

\begin{equation}
    C=\frac{1}{A_{0}^{a+b+1}B(a+1,b+1)},
\end{equation}

\noindent where $B(x,y)=\int_0^{1} t^{x-1}(1-t)^{y-1}dt$ is the beta function such that we have $\int_{0}^{A_0}P(A)dA=1$.

\begin{figure*}[ht]
    \includegraphics[width=\linewidth]{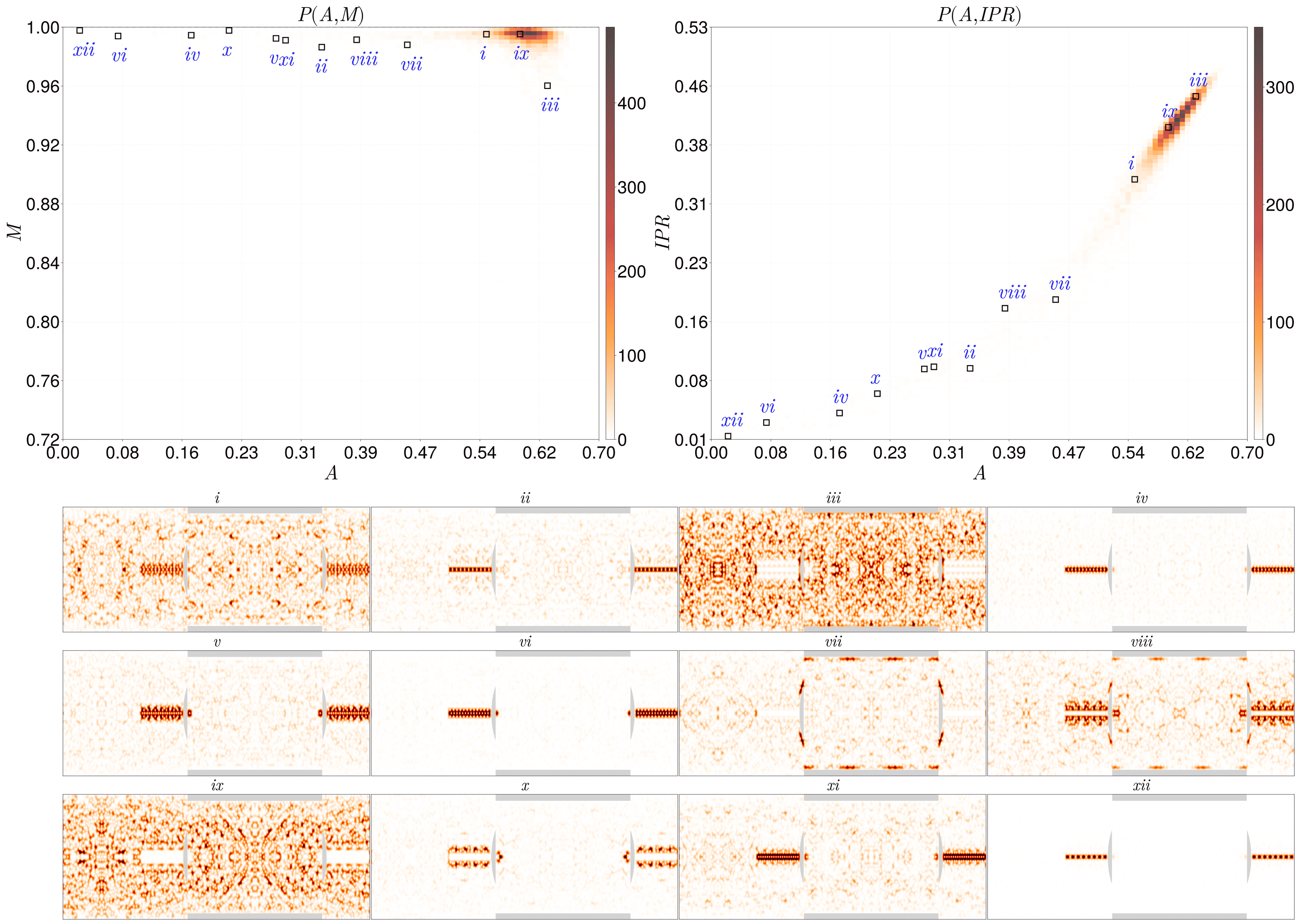}
    \put(-0.92\linewidth,0.4\linewidth){\textbf{(a)}}
    \put(-0.1\linewidth,0.4\linewidth){\textbf{(b)}}
    \caption{Joint probability density $P(A,M)$ (a) and $P(A,IPR)$ (b)  for chaotic states for stem width $w=0.9$ in the wavenumber window centered at $k \approx 2000.0$ containing approximately $10000$ PH functions in total (here showing only chaotic). (Below) Some representatives from the PH function ensemble with small $A$ highlighting the most strongly localized states that produce long tails in the numerical distributions. The gray area in PH function plots is the regular component. The scale in both heatmaps is linear, signifying counts of PH functions in a \([A,A+\delta A]\times[M,M+\delta M]\) or \([A,A+\delta A]\times[\mathrm{IPR},\mathrm{IPR}+\delta\mathrm{IPR}]\) bin, respectively.}
    \label{combined_w_0.9}
\end{figure*}

\noindent In the semiclassical limit PUSC implies that PH functions must become fully extended on the chaotic quantum phase space component with the distribution of entropy localization measures nearing the Dirac delta distribution $P(A) \rightarrow \delta (A - A_{max})$. The value of $A_{max} \approx 0.66$ is due to the oscillatory nature of the PH function \cite{BLR2019B} and is close to the value observed in many other systems \cite{LLR2021,BLR2020}.

Here we study the statistics of localization measures for chaotic states. The dependence of localization on stem half-width $w$ is striking, with narrow and wide stems exhibiting qualitatively different behavior even at very large wavenumbers.  In Fig.\ref{localizations_w_0.1} we plot the histogram of entropy localization measures $A_n$ for $w=0.1$ in sliding windows of $2000$ eigenstates, centered at increasingly large k up to $k\approx 4\,000$.  Despite this high-energy regime, the empirical distribution remains broad and far from the Dirac delta peak expected in the far semiclassical limit.  In particular, the tails of the distribution extend deeply into the small-A region, signaling a substantial fraction of eigenstates that remain tightly localized in the bouncing-ball region of phase space.  Fitting this curve with the beta distribution \eqref{beta_distribution} shows the low-A tail, meaning that dynamical localization remains strong for such narrow stems.

Moving to an intermediate stem width, $w=0.6$, Fig.\ref{localizations_w_0.6} shows that the entropy localization measure distribution quickly approaches the beta distribution as k increases.  At lower $k$, one still sees a small excess of small-$A$ events; but this deviation steadily shrinks as the window slides to higher energies.  By $k\approx 2200$ the fit is excellent across nearly the entire support, demonstrating that moderate widening of the stem greatly accelerates the onset of quantum ergodicity in the chaotic component.

Finally, in Fig.\ref{localizations_w_0.9} we examine $w=0.9$, the widest stem considered. Even at relatively small k the distribution of $A_n$ becomes sharply peaked around its maximum value, and by $k\approx 2000$ the histogram is essentially indistinguishable from a very narrow beta distribution.  The remaining small deviations in the tail correspond to rare bouncing-ball–scarred modes, whose proportion is vanishingly small in this geometry.

\begin{figure}[!h]
  \begin{subfigure}[t]{0.5\linewidth}
    \begin{overpic}[width=\linewidth,height=1.8in]{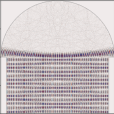}
      \put(3,90){\large (a)}
    \end{overpic}
  \end{subfigure}\hfill
  \begin{subfigure}[t]{0.5\linewidth}
    \begin{overpic}[width=\linewidth,height=1.8in]{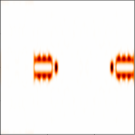}
      \put(1,89){\large\ (b)}
    \end{overpic}
  \end{subfigure}
  \caption{
    Example of a near bouncing-ball mode heavily scarred state ($k=369.205$) with stem width $w=0.9$.}
    \label{near_BB}
\end{figure}

\begin{figure*}
    \includegraphics[width=1.0\linewidth,height=6.5in]{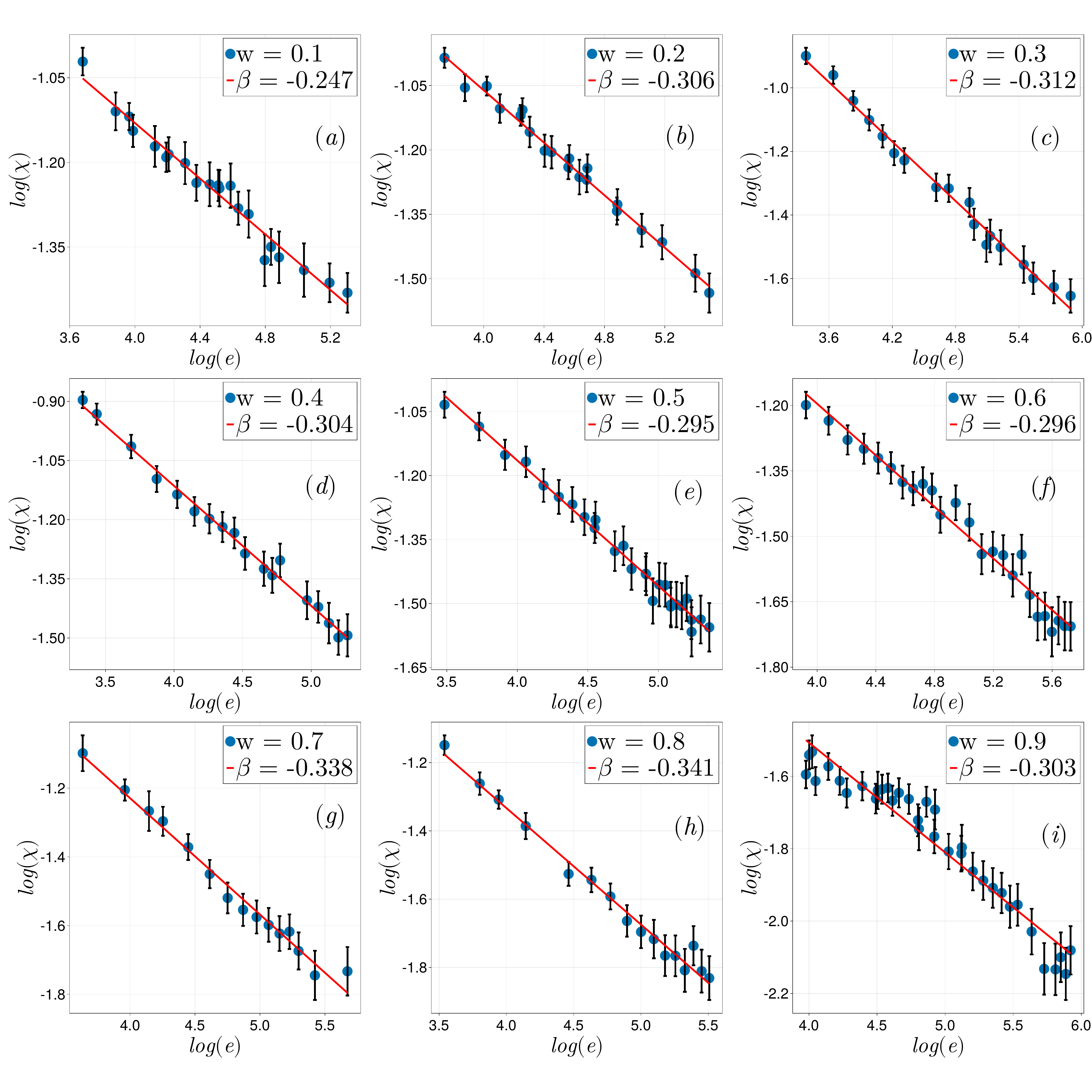}
    \caption{Log-log plots of the fraction of mixed states $\chi$ as a function of the unfolded energy $e$ for states with $M \in (-0.8,0.8)$. The error bars show $\pm 1 \sigma$ deviation from the mean. The fraction of mixed states was calculated using \eqref{mixed_states_eq}. A clear linear fit is observed over two decadic orders of magnitude for all stem half-widths $w=0.1-0.9$.}
    \label{mixed_states_main}
\end{figure*}

To elucidate the relationship between entropy localization measure and inverse participation ratio, we turn to the joint densities $P(A,M)$ and $P(\mathrm{IPR},A)$ for the cases $w=0.1$,$w=0.6$ and $w=0.9$. Fig.\ref{combined_w_0.1}(a) shows $P(A,M)$ for $w=0.1$, in which the low-A region is heavily populated, reflecting strong localization on the bouncing‐ball region deep within the chaotic sea.  Fig.\ref{combined_w_0.1}(b) reveals a nearly linear correlation between $A_n$ and $\mathrm{IPR}_n$ across most of the ensemble, with only minor scatter at very small $A$. Fig.\ref{combined_w_0.6}(a) shows the joint probability density results for $w=0.6$ showing a more narrow spread of $M$ indexes of PH functions, mostly confined to the large $A$ region. This is reflected in the relationship between $IPR_n$ and $A_n$ where a tail of highly localized states remains, as seen in Fig.\ref{combined_w_0.6}(b). Finally, the case of $w=0.9$ (Fig.\ref{combined_w_0.9}(a)) demonstrates that the joint distribution collapses tightly around large $A$ and $M$, with only a faint remnant tail due to isolated bouncing-ball modes—whose relative frequency itself falls off as a power-law in k \cite{Backer_1997}.  Indeed, these exceptional states are directly visualized in Fig.\ref{near_BB}, where panels (a) and (b) display, respectively, the configuration-space and PH images of a near–bouncing-ball region for $w=0.9$ as seen in Fig.\ref{combined_w_0.9}.  In all, these comparisons underscore how stem width controls the strength and persistence of dynamical localization in the mushroom. Nevertheless, localization effects are minor for $w \ge 0.3$, as shown by the spectral statistics reported in our previous work \cite{OreLozRobYan2025}.

\section{Decay of mixed states}
\label{mixed_states}

According to PUSC, the fraction of mixed eigenstates must vanish in the strict semiclassical limit.  However, there is no general theory governing the rate at which this fraction decays as one increases the effective semiclassical parameter (effective Planck constant). Consequently, the numerical determination of this decay law constitutes a central result of the present work.  We quantify the mixed–state fraction by

\begin{equation} \label{mixed_states_eq}
    \chi(e)=\frac{N(M_{th}^- \le M \le M_{th}^+)}{N},
\end{equation} 

\noindent where the numerator counts the mixed-type eigenstates within an energy window centered on e \eqref{semiclassical_parameter}, and $M_{\rm th}^\pm$ are the thresholds used to classify states via their M–index, inside an energy window containing $N$ eigenstates.

\begin{table*}
  \small
  \setlength{\tabcolsep}{6pt}
  \begin{tabular*}{\textwidth}{@{\extracolsep{\fill}} l c c c c c @{}}
    \toprule
    & \(-0.5\le M\le0.5\) & \(-0.6\le M\le0.6\) & \(-0.7\le M\le0.7\) & \(-0.8\le M\le0.8\) & \(-0.9\le M\le0.9\) \\
    \midrule
    \(w=0.1\) & (-0.260,0.950) & (-0.265,0.982) & (-0.273,0.977) & (-0.247,0.971) & (-0.273,0.975) \\
    \(w=0.2\) & (-0.287,0.960) & (-0.265,0.941) & (-0.275,0.957) & (-0.306,0.989) & (-0.288,0.962) \\
    \(w=0.3\) & (-0.306,0.981) & (-0.296,0.988) & (-0.278,0.985) & (-0.312,0.989) & (-0.338,0.997) \\
    \(w=0.4\) & (-0.322,0.968) & (-0.297,0.981) & (-0.264,0.977) & (-0.304,0.991) & (-0.340,0.973) \\
    \(w=0.5\) & (-0.257,0.902) & (-0.257,0.948) & (-0.262,0.966) & (-0.295,0.984) & (-0.307,0.986) \\
    \(w=0.6\) & (-0.369,0.954) & (-0.355,0.962) & (-0.295,0.960) & (-0.296,0.971) & (-0.276,0.987) \\
    \(w=0.7\) & (-0.337,0.980) & (-0.334,0.965) & (-0.292,0.970) & (-0.338,0.983) & (-0.329,0.985) \\
    \(w=0.8\) & (-0.307,0.970) & (-0.310,0.970) & (-0.305,0.963) & (-0.341,0.985) & (-0.347,0.990) \\
    \(w=0.9\) & (-0.314,0.900) & (-0.357,0.943) & (-0.321,0.940) & (-0.303,0.930) & (-0.279,0.810) \\
    \bottomrule
  \end{tabular*}
  \caption{Table containing the $(\beta,R^2)$ pairs for each threshold criteria $(-M_{th}^-\le M\le M_{th}^+)$ for all geometries in our study. $\beta$ is the power-law exponent and $R^2$ is the coefficient of determination for the log-log linear fit. A smaller value of $R^2$ is systematically seen in the case of stem width $w=0.9$. The determination of mixed type states is rather difficult in that geometry due ro the smallness of the regular phase space component and the broader coherent state peak in the PH function.}
  \label{mixed_data_table}
\end{table*}

Fig.\ref{mixed_states_main} shows our numerical results in the form of a log-log plot of the fraction of mixed states versus the semiclassical parameter $e$ \eqref{semiclassical_parameter}, where a clear power‐law decay

\begin{equation}
     \chi(e) \propto e^{\beta},
\end{equation}

\begin{figure}
    \includegraphics[width=1.0\linewidth]{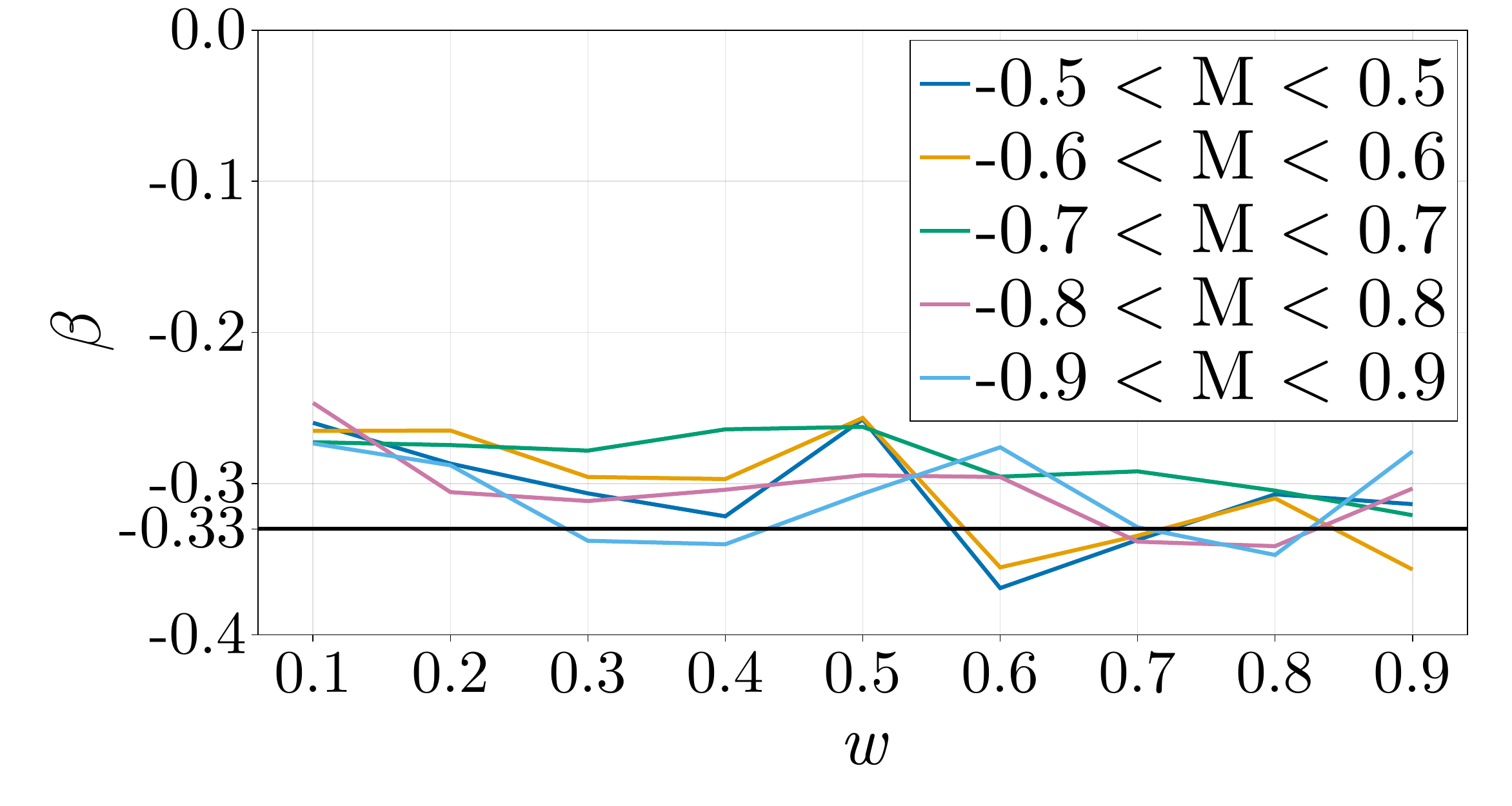}
    \caption{Decay parameter $\beta$ as a function of stem width $w$ for various symmetric thresholding criteria. It can be seen that all geometries have a decay exponent $\beta \approx -\frac{1}{3}$ independent of the specific threshold criteria.}
    \label{beta_vs_thresh}
\end{figure}

\noindent is observed with fitted exponents $\beta\simeq -1/3$ across all stem widths. In fact we observe a trend of slightly larger $\beta$s for smaller stem widths $w$ (Fig.\ref{beta_vs_thresh}) across different choices of $(M_{th}^-,M_{th}^+)$, where an agreement with the  power-law decay was observed for all threshold choices, across all stem widths (Fig.\ref{beta_vs_thresh}). Table \ref{mixed_data_table} compiles the fitted $\beta$ exponents together with their coefficients of determination ($R^2$), demonstrating that $\beta$ remains stable across changes in $(M_{\mathrm{th}}^{-}, M_{\mathrm{th}}^{+})$. This was necessary to explore because the coherent states used to build the PH function have finite width, causing some regular eigenstates localized on the outermost invariant tori to "leak" across the boundary between the chaotic and regular region. As a result these modes can be spuriously classified as mixed-type states. Although the relative fraction of such outliers is already very small, narrowing the acceptance window $(M_\mathrm{th}^-,M_\mathrm{th}^+)$ further suppresses them and ensures a more accurate count of genuine mixed-type states. For most widths and threshold choices, $R^2$ exceeds 0.95, reflecting an excellent linear fit in the log–log plot; the weakest correlation ($R^2\approx0.81$) occurs for the widest stem w=0.9 under the broadest threshold $-0.9\le M\le 0.9$. Due to the smallest Liouville regular phase space volume the fraction of mixed-type states is small implying a larger numerical error.

\section{Summary, Conclusion and Discussion}
\label{conclusion}

In this work we have carried out a comprehensive investigation of phase‐space localization in chaotic eigenstates for a one‐parameter family of Bunimovich mushroom billiards, with stem half-width \(w\) varying from 0.1 to 0.9.  Building on our previous spectral analysis including the gap‐ratio statistics in the same geometries, we have focused here on the structure of individual eigenfunctions as revealed by their Poincar\'{e}-Husimi (PH) representations, and on the quantitative measures of localization they exhibit.

Our main findings can be grouped into two broad themes:

\begin{enumerate}
  \item \emph{Stem‐width dependence of localization.}  
    We have shown that narrow stems amplify bouncing‐ball stickiness deep in the chaotic region, leading to pronounced dynamical localization even at high wavenumbers \(k\).  Conversely, wider stems exhibit far more uniform spreading of chaotic eigenstates, in good agreement with the two‐parameter beta distribution often observed in fully ergodic billiards without stickiness regions.  

  \item \emph{Mixed‐type states power-law decay.}  
    By applying a separation criterion to our PH ensemble, we tracked the fraction of mixed eigenstates—those with phase‐space support straddling both regular and chaotic regions—across increasing semiclassical parameter $e$.  Across all \(w\), \(\chi(e)\) decays as a robust power-law, \(\chi\propto e^{\beta}\), with \(\beta\approx -\tfrac13\).  We also observe a systematic albeit mild trend of slower decay (\(|\beta|\) smaller) for the narrowest stems, consistent with their enhanced boundary stickiness.  
\end{enumerate}

\medskip

Our numerical results highlight a profound connection between classical stickiness mechanisms—such as the marginally unstable periodic orbits that lie along the cap–stem separatrix and the families of bouncing‐ball trajectories—and the emergence of quantum localization phenomena.  We observe a robust power‐law decay exponent $\beta \approx -1/3$, matching findings in other mixed‐type systems and suggesting the existence of a universality class governing the decay of mixed states \cite{YWR2024,LLR2022,Yan_FPUT_mixed,Wang_Dicke_mixed}.  At the same time, the slight dependence of $\beta$ on the stem width w indicates that specific geometric features imprint quantitative signatures on this decay, pointing to a rich interplay between global universality and system‐dependent corrections.  Deriving the exponent $\beta$ from first principles remains an important open problem. 

Our study exploits PH functions for their manifest positivity and their direct relation to boundary‐data representations of eigenstates (see Sec. \ref{PH}).  However, this approach necessarily inherits systematic effects from the finite wavenumber k and the discrete grid sampling we employ.  In particular, the resolution limits of the Poincaré section and our chosen coherent‐state width introduce a smoothing that can sometimes obscure the sharpest scarring features and the finest-scale structure of the border region between the chaotic and regular regions, as observed in the case of $w=0.9$.

The numerical results point to a necessity of further study. Potentially, a time‐dependent study of wave‐packet evolution in the same geometries could dynamically clarify how the Ehrenfest and Heisenberg time scales manifest, measuring directly when classical mixing gives way to quantum interference \cite{silvestrov_ehrenfest,Haake}.  Second, examining other mixed‐type phase space billiards—such as mushroom billiards with elliptic caps or triangular stems \cite{Dietz_mushroom}—would isolate the effects of boundary curvature and corner angles on stickiness and localization; notably, recent work shows that power‐law decay of mixed‐type phase space states can occur even without explicit stickiness at the boundary \cite{LLR2021}.  Finally, experimental realizations in microwave resonators or optical wave guide arrays shaped like mushroom billiards would allow direct imaging of PH‐like intensity patterns, bringing these theoretical insights into the laboratory \cite{Stoe}.

Finally, our results reinforce PUSC in that, at sufficiently high \(k\), chaotic eigenfunctions indeed fill the entire chaotic region uniformly.  Yet the persistence of mixed states at intermediate \(k\) underscores the richness of the approach to the far semiclassical limit, and the need for refined theoretical tools to capture pre‐asymptotic behavior in mixed‐type quantum systems.

\section*{Acknowledgements}

We thank Dr. Hua Yan for stimulating discussions and suggestions and Dr. Črt Lozej for additional comments. This work was supported by the Slovenian Research and Innovation Agency (ARIS) under grants J1-4387 and P1-0306, and made use of the HPC VEGA supercomputer system under project S24O02-01.

\bibliography{main_final.bbl}

\appendix

\section{Stickiness on the boundary}
\label{stickiness_detailed}

\begin{figure}
    \centering
    \includegraphics[width=\linewidth]{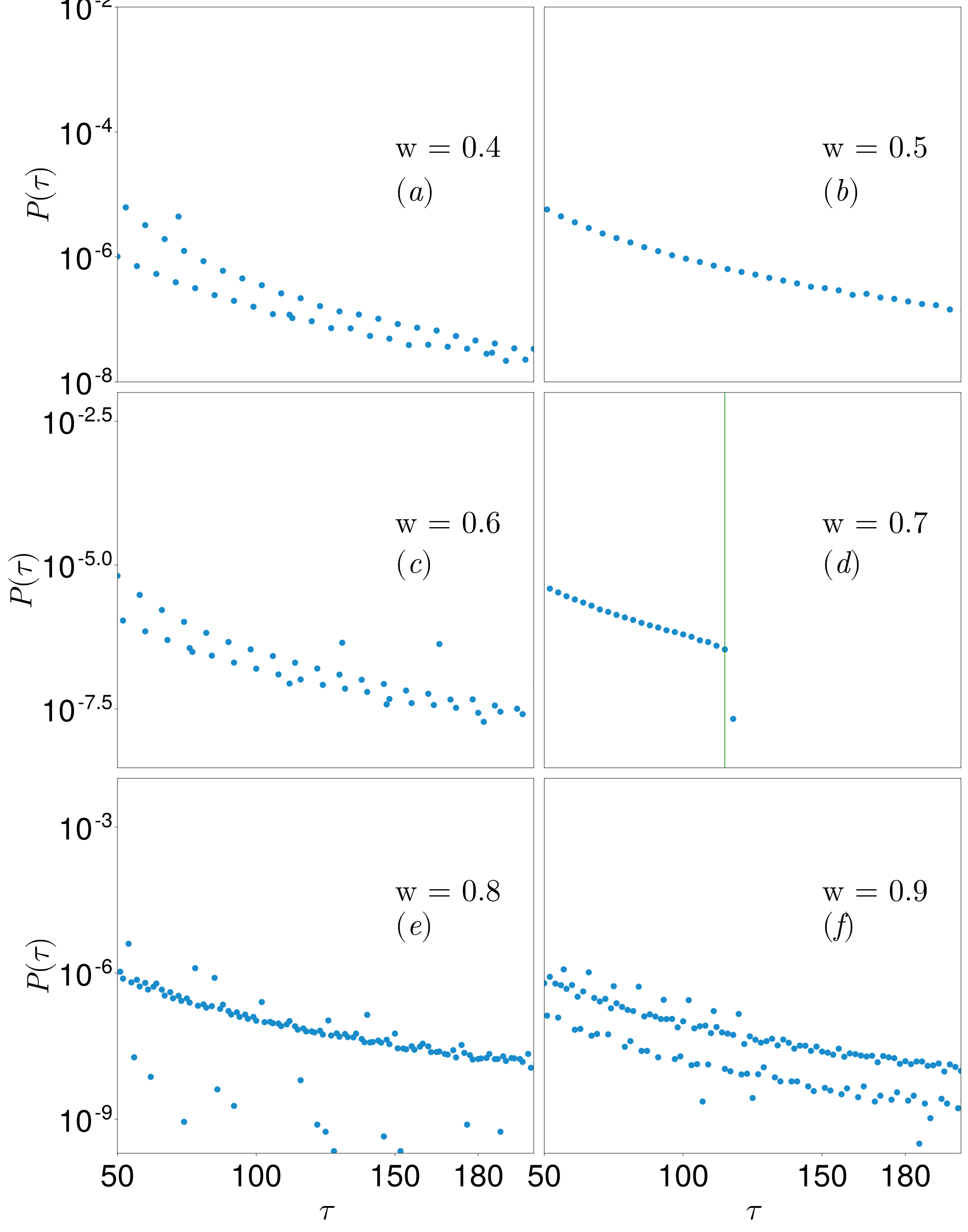}
    \caption{$P(\tau)$ for all stem half-widths $w$ of the triangle-stem mushroom billiard. If \eqref{MUPO_check} holds, periodic sequences of recurrence times \protect\cite{Altmann2005} appear for each MUPO found, showing as bands in the scatter plot. For w=0.7, the sequence breaks (green line), as a consequence of this geometry not admitting a MUPO. Therefore no power-law tail is present \protect\cite{Altmann2005}. The plot is shown in log-lin scale with no fit to the data. A single randomly selected initial condition within the chaotic component was used for each $w\in[0.4,0.9]$ and iterated $N=3\times10^{9}$ times. The recurrence statistics were found to be independent of the particular choice of initial condition within the chaotic component.}
    \label{P_tau}
\end{figure}

We first applied the continued-fraction expansion as detailed in Dettmann et al. \cite{dettman_continued_fraction} to test the cap MUPO criterion \eqref{MUPO_check}. For each half-width w we considered the geometry of a triangular stem mushroom billiard. We then used recurrence time statistics, choosing the recurrence region to isolate the relevant mechanism: given a region $A$, the recurrence time $\tau$ is the number of collisions outside $A$ between two successive visits to $A$ (recurrence into $A$). For the case of the triangular stem mushroom billiard we took A to be the stem (foot), so each recurrence time $\tau$ equals the number of cap collisions between successive visits to the stem. If the half-width $w$ satisfies \eqref{MUPO_check} (therefore we have a MUPO) then this choice of $A$ gives a periodic sequence of recurrence times $\tau$ as seen in Fig.\ref{P_tau} \cite{Altmann2005}. Checks for the MUPO criterion \eqref{MUPO_check} are shown in Table.\ref{tab:mupo_q_eta}. Plotting the probability density of recurrence times $P(\tau)$ (Fig.\ref{P_tau}) we observe that:

\begin{equation} \label{periodic_recurrence_eq}
    \tau = q + 2\eta
\end{equation}

\begin{figure}
    \centering
    \includegraphics[width=\linewidth]{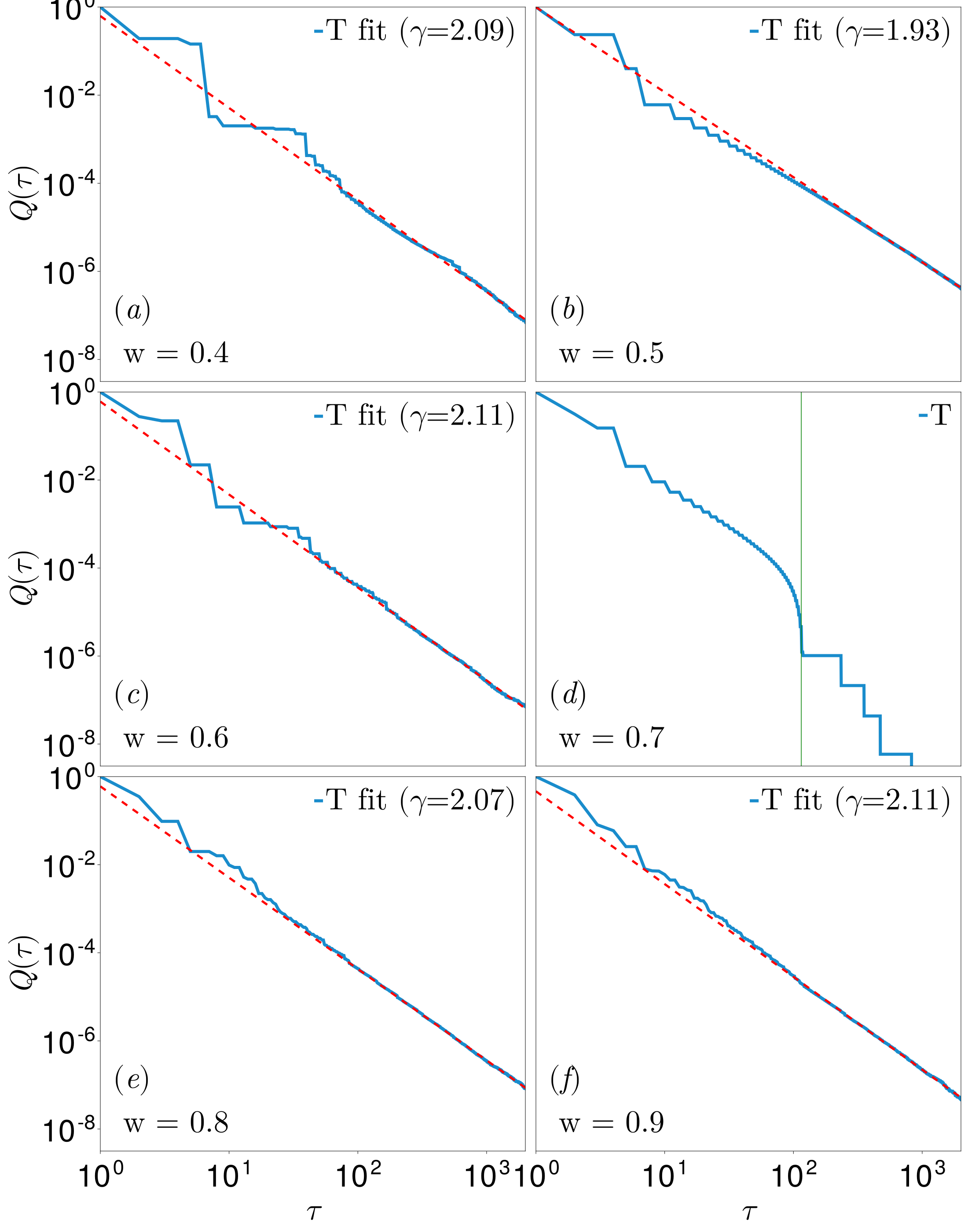}
    \caption{Log-log plot of $Q(\tau)$ for half-widths $w\ge0.4$ of the triangle stem mushroom $(T)$ billiard showing a good agreement with the expected power law exponent of $\gamma=2$. The fit was performed on the tail of the distribution and the result was extrapolated across the plotted range. The case of $w=0.7$ shows no power-law tail due to the fact that there is no MUPO associated with this particular geometry (see Table.\ref{tab:mupo_q_eta}).} 
    \label{Q_tau}
\end{figure}

\begin{table}[b]
\centering
\small
\begin{tabular*}{0.5\columnwidth}{@{\extracolsep{\fill}} c l}
\toprule
$w$ & $(q,\eta)$ \\
\midrule
0.4 & $(8,3),\ (84,31)$ \\
0.5 & $(3,1)$ \\
0.6 & $(10,3),\ (44,13)$ \\
0.7 & --- \\
0.8 & $(4,1)$ \\
0.9 & $(6,1)$ \\
\bottomrule
\end{tabular*}
\caption{Smallest MUPO candidates ($q<100$) for selected stem half-widths $w$, listed as $(q,\eta)$. Cases $w=0.1,0.2,0.3$ are not shown because they do not satisfy \eqref{MUPO_check}.}
\label{tab:mupo_q_eta}
\end{table}

\noindent where $\tau$ is the recurrence time and $(q,\eta)$ are MUPO indexes found in the continued fraction decomposition (if a MUPO is present, otherwise \eqref{periodic_recurrence_eq} does not hold). Another useful diagnostic of recurrence time analysis is also the survival time probability function:

\begin{equation}
    Q(\tau)=\sum_{T=\tau}^{\infty}P(T)=\lim_{N \rightarrow \infty} \frac{N_{\tau}}{N}
\end{equation}

\noindent defined as the sum of all recurrence time probability densities $P(T)$ whose recurrence time $T$ is larger than $\tau$ or, equivalently, in the limit of taking a very large number of collisions $N$ the fraction of recurrences $N_{\tau}$ with time $T\geq \tau$. If stickiness due to MUPOs is present it should be seen as a power-law in the tail of $Q(\tau)$:

\begin{equation} \label{Q_tau_power_law}
    Q(\tau) \propto \tau^{-\gamma}
\end{equation}

\noindent with an exponent $\gamma=2$. Stickiness associated with bouncing-ball modes in the stem is ubiquitous, whereas stickiness due to MUPOs in the cap need not occur, as seen in Fig.\ref{Q_tau}(d) for the case of $w=0.7$, as there is no power-law tail seen. This is also seen in Fig.\ref{P_tau}(d) where there is an abrupt stop in the sequence of recurrence times with small sequential differences (represented as a vertical green line). This is expected because this geometry (alongside $w=0.1,0.2,0.3$) has no associated cap MUPOs with $\eta<100$ (see Table.\ref{tab:mupo_q_eta}). Nevertheless, some residual cap stickiness persists, arising from stem orbits that intersect the inscribed dashed circle in Fig.~\ref{geom_mushroom}~\cite{Georgiou2011Thesis}.

\section{Lagrangian Descriptors (LD)}
\label{descriptors}

\begin{figure*}
  \begin{subfigure}[t]{0.5\linewidth}
    \begin{overpic}[width=\linewidth]{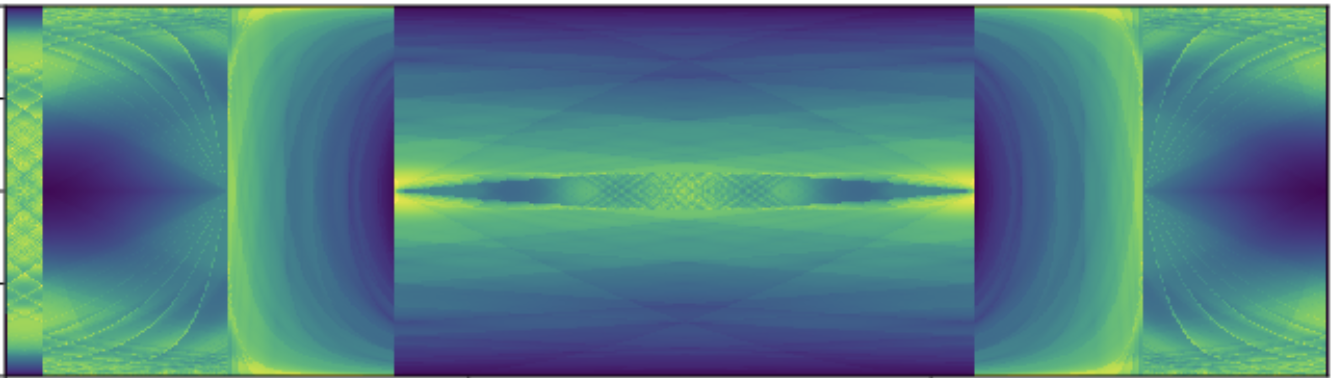}
      \put(1,22){\color{red}\large (a)}
    \end{overpic}
  \end{subfigure}\hfill
  \begin{subfigure}[t]{0.5\linewidth}
    \begin{overpic}[width=\linewidth]{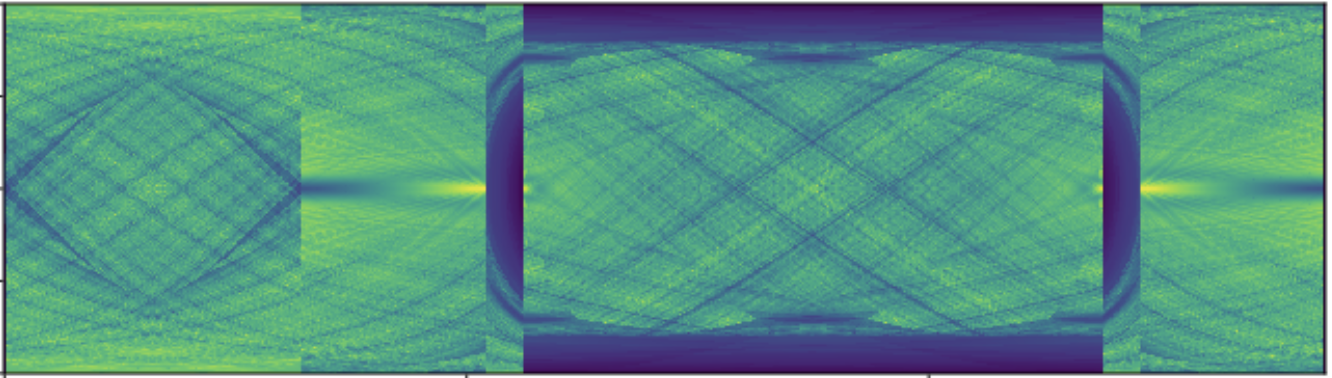}
      \put(1,22){\color{red}\large (b)}
    \end{overpic}
  \end{subfigure}
  
  \begin{subfigure}[t]{0.5\linewidth}
    \begin{overpic}[width=\linewidth]{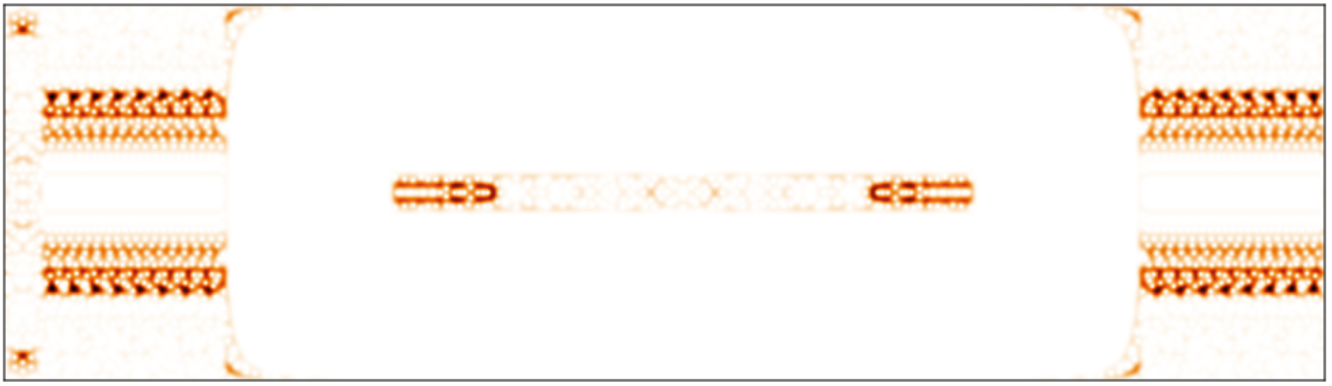}
      \put(1,22){\color{blue}\large (c)}
    \end{overpic}
  \end{subfigure}
  \hfill
  \begin{subfigure}[t]{0.5\linewidth}
    \begin{overpic}[width=\linewidth]{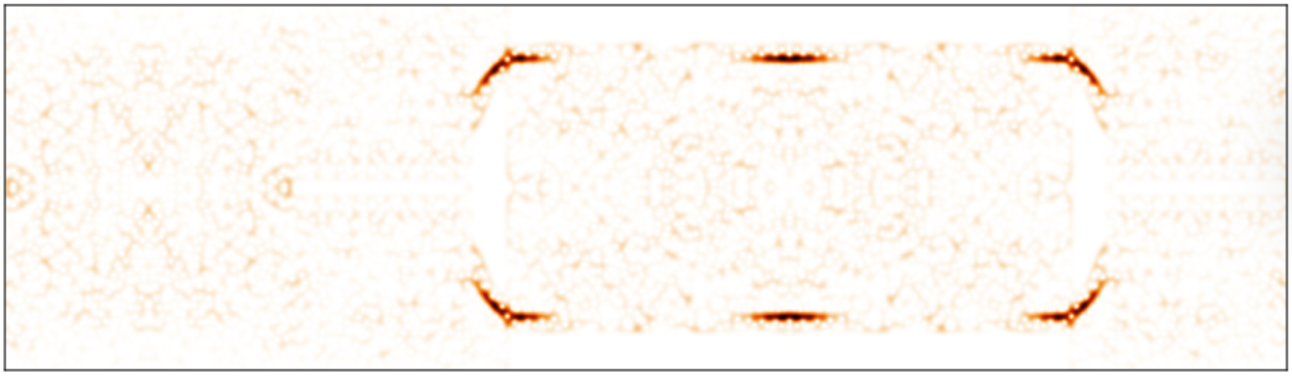}
      \put(1,22){\color{blue}\large (d)}
    \end{overpic}
  \end{subfigure}

   \begin{subfigure}[t]{0.5\linewidth}
    \begin{overpic}[width=\linewidth]{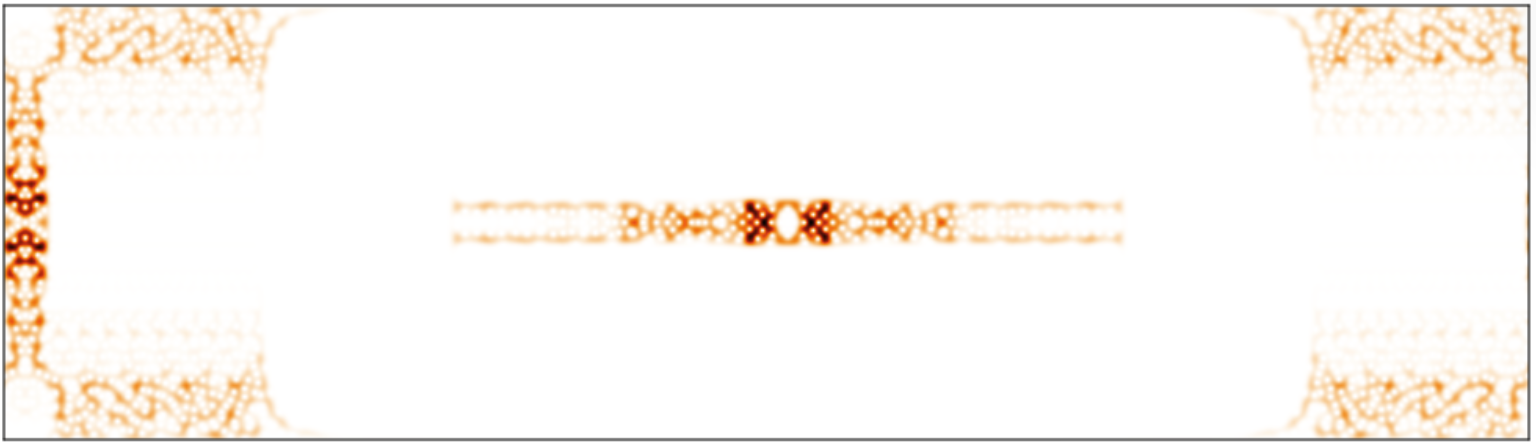}
      \put(1,22){\color{blue}\large (e)}
    \end{overpic}
  \end{subfigure}
  \hfill
  \begin{subfigure}[t]{0.5\linewidth}
    \begin{overpic}[width=\linewidth]{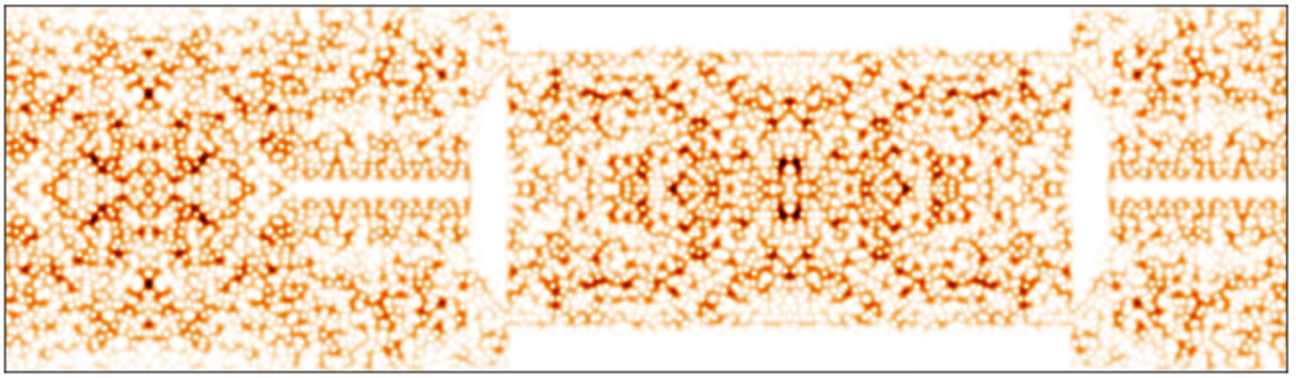}
      \put(1,22){\color{blue}\large (f)}
    \end{overpic}
  \end{subfigure}

  \caption{
    Comparison of classical phase space structures as seen by the LD for cases of $w=0.1$ (a) and $w=0.8$ (b). LD values span linearly from 0 to 1 with higher values appearing brighter. (c) ($k \approx 3000.076 $) and (e) $k \approx 3000.089 $) represent complementary PH function plots showing that classical phase structures influence PH function structure. (d) $(k \approx 3000.047) $ shows PH function localization on a sticky MUPO region inside the chaotic sea with $M=1$ \eqref{m_index} close to the separatrix in classical phase space that cannot be seen with the S plot recurrence analysis. (f) ($k \approx 3000.087 $) shows a complementary PH function to (d) showing decreased density around the previous aforementioned MUPO and the influence of the classical phase space structures on the patterns found in the stem and cap region of phase space.}
    \label{LD_PH_comparison}
\end{figure*}

For the representative geometries $w=0.1$ and $w=0.8$ we have studied the classical Lagrangian descriptors and their manifestation in some representative PH functions. Lagrangian descriptors provide a complementary way of visualizing the classical Poincar\'e surface of section (PSOS), showing structures related to classical transport (stable/unstable manifolds, MUPOs etc.). This appears as a sensitive change in the LD value as the initial condition (s,p) is varied. These calculations were performed for the rectangular-stem mushroom billiard. We have used the following construction of the LD \cite{lagrangian_descriptors}:

\begin{gather} \label{LD}
    LD = LD^{+} \times LD^{-} \\
    LD^{+} = \sum_{t=0}^{N_f} \left(|\Delta s_{t}|^a + |\Delta p_{t}|^a \right) \\
    LD^{-} = \sum_{t=-N_b}^{-1} \left(|\Delta s_{t}|^a + |\Delta p_{t}|^a \right) \\
    \Delta s_{t} = s_{t+1} - s_{t} \quad and \quad \Delta p_{t} = p_{t+1} - p_{t}
\end{gather}

\noindent where $LD^{+}$ and $LD^{-}$ denote the forward- and backward-time Lagrangian descriptors; $\Delta s_{t}$ and $\Delta p_{t}$ are the successive arc-length and momentum differences, respectively; and $N_f$ and $N_b$ are the numbers of forward and backward trajectory iterations, computed using the $L^{a}$ norm with $a=0.99$ (empirically chosen to maximize contrast in classical phase space structures). In our calculations we chose $N_f=N_b=10$ with $N=10^6$ initial conditions. In Fig.\ref{LD_PH_comparison} we compare the LD heatmap with PH functions corresponding to chaotic eigenstates for the cases of $w=0.1$ and $w=0.8$. It shows that classical phase space structures as seen in the LD are found in the patterns of the PH functions, showing the influence of classical transport in PH functions.

\end{document}